\documentclass[10pt]{article}

\usepackage{amsmath}

\usepackage{array}

\usepackage{appendix}

\usepackage{tocloft}                   

\usepackage{graphicx}

\usepackage{amsfonts}

\usepackage{amssymb}

\usepackage{mathrsfs}

\usepackage{yfonts}

\usepackage{euscript}

\usepackage{centernot}                 

\usepackage{ifsym}                     

\usepackage{lmodern}                   

\usepackage{upgreek}

\usepackage{mathtools}

\usepackage{bm}

\usepackage{undertilde}

\usepackage{amsfonts}

\usepackage{amssymb}

\usepackage{mathrsfs}

\usepackage{upgreek}

\usepackage{mathtools}

\usepackage{color}

\usepackage{slantsc}
\usepackage{calligra}

\usepackage{bbold}          

\usepackage[T1]{fontenc}

\usepackage{epsf}

\usepackage{latexsym}

\usepackage{tipa}

\usepackage{makeidx}

\makeindex

\def\b{\begin{equation}}

\def\e{\begin{equation}}

\def\be{\begin{equation}}              

\def\ee{\end{equation}}

\def\beq{\begin{equation}}

\def\eeq{\end{equation}}

\def\bea{\begin{eqnarray}}

\def\eea{\end{eqnarray}}

\def\m{\mbox{ }}

\def\mma {\m , \m \m }

\def\!{\hspace{-1.6667em}}

\def\n{\noindent}

\def\u{\underline}

\def\uc{\underbracket}   
\def\uo{\utilde}     

\def\bia{\mbox{\boldmath$a$}}

\def\bic{\mbox{\boldmath$c$}}

\def\bif{\mbox{\boldmath$f$}}

\def\bip{\mbox{\boldmath$p$}}

\def\bis{\mbox{\boldmath$s$}}

\def\biu{\mbox{\boldmath$u$}}

\def\biA{\mbox{\boldmath$A$}}

\def\biC{\mbox{\boldmath$C$}}              

\def\biD{\mbox{\boldmath$D$}}

\def\biF{\mbox{\boldmath $F$}}             %

\def\biP{\mbox{\boldmath$P$}}

\def\biQ{\mbox{\boldmath$Q$}}

\def\biR{\mbox{\boldmath$R$}}

\def\biS{\mbox{\boldmath$S$}}

\def\biT{\mbox{\boldmath$T$}}

\def\biU{\mbox{\boldmath$U$}}

\def\biV{\mbox{\boldmath$V$}}

\def\biW{\mbox{\boldmath$W$}}

\def\sbiU{\mbox{\ttfamily\fontseries{b}\selectfont U}}                   %

\def\sbiC{\mbox{\ttfamily\fontseries{b}\selectfont C}}

\def\sbiK{\mbox{\ttfamily\fontseries{b}\selectfont K}}

\def\sbiA{\mbox{\ttfamily\fontseries{b}\selectfont A}}

\def\sbiD{\mbox{\ttfamily\fontseries{b}\selectfont D}} 

\def\sbiu{\mbox{\scriptsize\boldmath$u$}}

\def\sbiN{\mbox{\scriptsize\boldmath$N$}}

\def\sbip{\mbox{\scriptsize\boldmath$p$}}

\def\sbupxi{\mbox{\scriptsize\boldmath$\upxi$}}

\def\mM{\mbox{M}}                        

\def\mN{\mbox{N}}

\def\ma{\mbox{a}}

\def\me{\mbox{e}}

\def\mg{\mbox{g}}

\def\mh{\mbox{h}}

\def\mi{\mbox{i}}

\def\ml{\mbox{l}}

\def\mn{\mbox{n}}  

\def\mo{\mbox{o}}

\def\mp{\mbox{p}}

\def\mr{\mbox{r}}

\def\ms{\mbox{s}}

\def\muu{\mbox{u}}

\def\bh{\u{\u{\mbox{h}}}  }            

\def\bG{\mbox{\bf G}}                     

\def\bM{\mbox{\bf M}}

\def\bN{\mbox{\bf N}}

\def\bX{\mbox{\bf X}}

\def\bY{\mbox{\bf Y}}

\def\bZ{\mbox{\bf Z}}

\def\bh{\mbox{\bf h}}

\def\bp{\mbox{\bf p}}

\def\bupSigma{\mbox{\boldmath$\Sigma$}}                 
\def\bupxi{\mbox{\boldmath$\xi$}}                       

\def\bupchi{\mbox{\boldmath$\chi$}}                     

\def\bcalD{\mbox{\boldmath ${\cal D}$}}

\def\fE{\mbox{\sffamily E}}

\def\fF{\mbox{\sffamily F}}

\def\fH{\mbox{\sffamily H}}   

\def\fI{\mbox{\sffamily I}}

\def\fJ{\mbox{\sffamily J}}

\def\fP{\mbox{\sffamily P}}

\def\fQ{\mbox{\sffamily Q}}

\def\fR{\mbox{\sffamily R}}

\def\fZ{\mbox{\sffamily Z}}

\def\sa{\mbox{\scriptsize a}}

\def\si{\mbox{\scriptsize i}}

\def\sll{\mbox{\scriptsize l}}  

\def\sm{\mbox{\scriptsize m}}

\def\sn{\mbox{\scriptsize n}}

\def\sr{\mbox{\scriptsize r}}

\def\st{\mbox{\scriptsize t}}

\def\sE{\mbox{\scriptsize E}}

\def\sS{\mbox{\scriptsize S}}

\def\sT{\mbox{\scriptsize T}}

\def\sfZ{\mbox{\sffamily{\scriptsize Z}}}      

\def\sbcC{\mbox{\boldmath \scriptsize ${\cal C}$}}

\def\sbcF{\mbox{\boldmath \scriptsize ${\cal F}$}}

\def\sbcG{\mbox{\boldmath \scriptsize ${\cal G}$}}

\def\sbcL{\mbox{\boldmath \scriptsize ${\cal L}$}}

\def\sbcS{\mbox{\boldmath \scriptsize ${\cal S}$}}

\def\bscC{\mbox{{\boldmath \scriptsize${\cal C}$}}}                               

\def\bscZ{\mbox{{\boldmath \scriptsize${\cal Z}$}}}                               

\def\bscF{\mbox{{\boldmath \scriptsize${\cal F}$}}}                               

\def\bscP{\mbox{\boldmath\scriptsize${\cal P}$}}                               

\def\btcP{\mbox{\boldmath\tiny${\cal P}$}}                                    

\def\btcQ{\mbox{\boldmath\tiny${\cal Q}$}}                                    

\def\bscS{\mbox{\boldmath \scriptsize${\cal S}$}}                               

\def\bscI{\mbox{{\boldmath \scriptsize${\cal I}$}}}                               

\def\bscM{\mbox{{\boldmath \scriptsize${\cal M}$}}}  

\def\bigupsigma{\mbox{\Large$\sigma$}}

\def\bttu{\mbox{\boldmath {\tt U}}}

\def\bttf{\mbox{\boldmath {\tt F}}}

\def\btts{\mbox{\boldmath {\tt s}}}

\def\Thomas{\,\,\mbox{\textcircled{$\rightarrow$}}\,\,}

\def\LThomas{\,\,\mbox{\textcircled{$\leftarrow$}}\,\,}

\def\TwoWay{\,\,\mbox{\textcircled{$\leftrightarrow$}}\,\,}

\def\sumi2{\sum\mbox{}_{\mbox{}_{\mbox{\scriptsize $i$=1}}}^2}

\def\sumi3{\sum\mbox{}_{\mbox{}_{\mbox{\scriptsize $i$=1}}}^3}

\def\sumABcycles3{\sum\mbox{}_{\mbox{}_{\mbox{\scriptsize cycles $A,B$=1}}}^{3}}

\def\sumCDcycles3{\sum\mbox{}_{\mbox{}_{\mbox{\scriptsize cycles $C,D$=1}}}^{3}}

\def\sumj3{\sum\mbox{}_{\mbox{}_{\mbox{\scriptsize $j$=1}}}^3}

\def\sumk3{\sum\mbox{}_{\mbox{}_{\mbox{\scriptsize $k$=1}}}^3}

\def\prodiA1{\prod\mbox{}_{\mbox{}_{\mbox{\scriptsize $i$=1}}}^{A - 1}}

\def\d{\textrm{d}}                                                  

\def\pa{\partial}                                                   

\def\bpa{\mbox{\boldmath$\partial$}}                                                   

\def\es{\m = \m}

\def\peq{\m \mbox{`='} \m}

\def\peqs{\m \m \mbox{`='} \m \m}

\def\:={\m := \m}

\def\=:{\m =: \m}

\def\FrA{\mbox{$\mathfrak{A}$}}                                

\def\FrC{\mbox{$\mathfrak{C}$}}                                
                                                       
\def\FrS{\mbox{\Large $\mathfrak{s}$}}                         
\def\lFrS{\mbox{\LARGE$\mathfrak{s}$}}                         

\def\lFrs{\mathfrak{S}}                                        

\def\FrU{\mbox{$\mathfrak{U}$}}                                
\def\FrD{\mbox{$\mathfrak{D}$}}	                               
 
\def\Frm{\mbox{\Large $\mathfrak{m}$}}                         
\def\FrM{\mbox{$\mathfrak{M}$}}                                
\def\lFrg{\mbox{\Large$\mathfrak{g}$}}                         
\def\FrK{\mathfrak{K}}                                         
\def\FrF{\mbox{\boldmath$\mathfrak{F}$}}                       

\def\sFrf{\mbox{\large $\mathfrak{f}$}}                          

\def\FrG{\mathfrak{G}}                                         
\def\sFrG{\mbox{\boldmath\scriptsize$\mathfrak{G}$}}           

\def\Hilb{\mbox{{\boldmath$\mathfrak{H}$}ilb}}                 

\def\scC{\mbox{\scriptsize ${\cal C}$}}                    

\def\btcC{\mbox{\tiny\boldmath ${\cal C}$}}                %

\def\scE{\mbox{\scriptsize ${\cal E}$}}                    

\def\scF{\mbox{\scriptsize ${\cal F}$}}

\def\scH{\mbox{\scriptsize ${\cal H}$}}                    

\def\bscP{\mbox{\boldmath\scriptsize ${\cal P}$}}

\def\bscP{\mbox{\boldmath\scriptsize ${\cal P}$}}

\def\bscR{\mbox{\boldmath\scriptsize ${\cal R}$}}

\def\bscQ{\mbox{\boldmath\scriptsize ${\cal Q}$}}

\def\bsG{\mbox{\boldmath\scriptsize G}}

\def\bscS{\mbox{\boldmath\scriptsize ${\cal S}$}}

\def\scS{\mbox{\scriptsize ${\cal S}$}}                    

\def\Sec{\bscS\mbox{\bf e}}                                

\def\bFlin{\sbcF\mbox{\bf lin}} 

\def\Chronos{\scC\mbox{hronos}}                            

\def\bGauge{\sbcG\mbox{\bf auge}} 

\def\bShuffle{\sbcS\mbox{\bf huffle}} 

\def\sFrO{\mbox{\boldmath\scriptsize$\mathfrak{O}$}}                        

\def\chronos{\mbox{\ttfamily\fontseries{b}\selectfont C}} 
					   
\def\biD{\mbox{\ttfamily\fontseries{b}\selectfont D}}      %
														   														   
\def\Dirac{\mbox{\ttfamily\fontseries{b}\selectfont D}}    
														   
\def\gauge{\mbox{\ttfamily\fontseries{b}\selectfont G}}                  

\def\Kuchar{\mbox{\ttfamily\fontseries{b}\selectfont K}}                  
\def\unres{\mbox{\ttfamily\fontseries{b}\selectfont U}}                   %

\def\unres{\mbox{\ttfamily\fontseries{b}\selectfont U}}
                                 
\def\FrQ{\mbox{\Large $\mathfrak{q}$}}                               
\def\bFrF{\mbox{\boldmath$\mathfrak{F}$}}                            %

\def\bFrG{\mbox{\boldmath$\mathfrak{G}$}}                            %
												  
\def\bFrC{\mbox{\boldmath$\mathfrak{C}$}}                            
\def\Phase{\mbox{{\boldmath$\mathfrak{P}$}hase}}                     

\def\bFrR{\mbox{\boldmath$\mathfrak{R}$}}                            
\def\Rig-Phase{\bFrR\mbox{ig-}\Phase}                                
                                                                													   
\def\sbiO{\mbox{\ttfamily \boldmath$O$}}                             

\def\sbiK{\mbox{\ttfamily \boldmath$K$}}                             

\def\sFrO{\mbox{\boldmath\scriptsize$\mathfrak{O}$}}     

\def\Obs{\FrO\mbox{bs}}                             %

\def\UnresObs{\FrU\mbox{nres-}\FrO\mbox{bs}}                                 %

\def\DiffObs{\FrD\mbox{iff-}\FrO\mbox{bs}}                                 %

\def\DiracObs{\FrD\mbox{irac-}\FrO\mbox{bs}}                                  %

\def\KucharObs{\FrK\mbox{uchar-}\FrO\mbox{bs}}                                  %

\def\GaugeObs{\FrG\mbox{auge-}\FrO\mbox{bs}}                                %

\def\ChronosObs{\FrC\mbox{hronos-}\FrO\mbox{bs}}                             %

\def\AObs{\FrA\mbox{-}\FrO\mbox{bs}}                                        %

\def\btsC{\mbox{\boldmath\tiny${\cal C}$}}                        

\def\btiO{\mbox{\boldmath\tiny$O$}}                        

\def\bFrZ{\mbox{\boldmath$\mathfrak{Z}$}}                            

\def\bFrR{\mbox{\boldmath$\mathfrak{R}$}}                            

\def\bFrR{\mbox{\boldmath$\mathfrak{R}$}}                            

\def\1mat{\u{\u{1}}}                                                 

\def\Positive-Modespace{\mbox{{\boldmath$\mathfrak{M}$}odespace$^+$}}

\def\POSITIVE-MODESPACE{\mbox{{\boldmath$\mathfrak{M}$}ODESPACE$^+$}}
                                                                                                                             														
\def\bFrG{\mbox{ $\mathfrak{G}$}}                                    %
\def\Riem{\bFrR\mbox{iem}}                                           

\def\FrO{\mbox{$\mathfrak{O}$}}                                      

\def\lattice{\mbox{\bf\Large$\mathfrak{L}$}}                                      

\def\sFrC{\mbox{\boldmath\scriptsize$\mathfrak{C}$}}                        

\def\Kin-Hilb{\mbox{{\boldmath$\mathfrak{K}$}in-\Hilb}}                     

\def\Mid-Hilb{\mbox{{\boldmath$\mathfrak{M}$}id-\Hilb}}                     

\def\Dyn-Hilb{\mbox{{\boldmath$\mathfrak{D}$}yn-\Hilb}}                     

\def\5Star{\mbox{\Large$\star$}}              

\def\K{Kucha\v{r} }

\def\Ks{Kucha\v{r}'s }

\def\peq{\m \mbox{`='} \m}

\def\speq{\m \peq \m}

\begin{document}

\begin{center}

\Huge{\bf A LOCAL RESOLUTION OF}

\Huge{\bf THE PROBLEM OF TIME}

\normalsize

\vspace{.1in}

\Large{\bf III. The other classical facets piecemeal}

\vspace{.1in}

{\large \bf E.  Anderson}$^1$

\vspace{.1in}

{\large \bf \it based on calculations done at Peterhouse, Cambridge} 

\end{center}

\begin{abstract}

We introduce nine further local facets of the classical Problem of Time, and underlying Background Independence aspects, 
the previous two Articles having covered one further facet-and-aspect each for a total of eleven. 
I.e.\ 1)  Constraint Closure, 
      2)  Assignment of Observables, 
      3)  Space Construction from Less Structured Space, 
	  A)  Spacetime Construction from Space,   
	  0') Spacetime Relationalism, 
	  1') Spacetime Generator Closure, 
	  2') Assignment of Spacetime Observables, 
      3') Spacetime Construction from Less Structured Spacetime, 	 
and   B)  Foliation Independence. 
These are classically implemented piecemeal by the Dirac Algorithm for 1) and the Lie Algorithm for 1'). 
Taking Function Spaces Thereover: over phase space for 2) and over the space of spacetimes for 2'). 
Feeding Deformed Families into the Dirac Algorithm for A) and into the Lie Algorithm for 3) and 3'), working out whenever Rigidity is encountered.  
Spacetime's diffeomorphism-invariance and perturbative Lie derivative for 0'). 
Finally, Refoliation Invariance following from the Dirac algebroid of the GR constraints for B).   
Each of these approaches is moreover grounded in Lie's Mathematics in accord with this Series' claim. 
The generalizations required to extend these resolutions to solve combined rather than just piecemeal facets is the main subject of Articles V to XIII.   
We include novel work on, firstly, Closure, including a blockwise breakdown into more specific Closure Problems. 
Secondly, on a coordinate-free Multi-tensor Calculus suitable for joint treatment of Background Independence aspects, 
presenting in particular the Dirac Algorithm, and ensuing theory of observables and of deformations. 
Finally, aspect 3') is new to the current Article. 

\end{abstract}

$^1$ dr.e.anderson.maths.physics *at* protonmail.com

\section{Introduction}

In this third Article on the Problem of Time \cite{Battelle, DeWitt67, Dirac, K81, K91, K92, I93, K99, APoT, FileR, APoT2, AObs, APoT3, ALett, ABook, A-CBI}
we introduce nine further local facets with their underlying Background Independence aspects, 
to Article I's \cite{I} Temporal Relationalism and Article II's \cite{II} Configurational Relationalism. 
For now, we work at the classical level, leaving discussion of the quantum counterparts of this total of nine local facets to Article IV \cite{IV}.  
Each of Temporal and Configurational Relationalism moreover provides constraint equations.
In the case of GR, these are the Hamiltonian and momentum constraints respectively.  

\m 

\n The current Article's nine aspects are as follows, 
beginning with the natural follow-up question of whether the constraints a theory obtains from Temporal and Configurational Relationalism form a complete and consistent picture.

\m 

%
\n{\bf Aspect 1: Constraint Closure} (in Sec 2 and Article VII)   
Ab initio this involves forming Poisson brackets between all constraints. 
If in the process second-class constraints are discovered, these brackets are to be replaced with Dirac bracket absorbing these second-class constraints.  

\m 

\n The {\bf Dirac Algorithm} \cite{Dirac, HTBook, ABook}(strategy 1) serves to reveal whether a set of constraints is inconsistent or incomplete. 
Constraint Closure thus has the status of a {\it necessary test for candidate} Temporally and Configurationally Relational theories, 
which fails if inconsistency arises. 
Incompleteness may moreover amount to a given group $\lFrg$, or group action thereof, being rejected by a candidate theory's configuration space $\FrQ$, 
or a given Principles of Dynamics action thereover being rejected.

\m 

\n{\bf The Constraint Closure Problem} (facet 1) refers to these possibilities of inconsistency or incompleteness. 

\m 

\n Successfully passing this test produces a Lie brackets algebraic structure of solely first-class constraints, 
with Poisson (or more generally Dirac) brackets in the role of Lie brackets.  
Relational Particle Mechanics (RPM) and minisuperspace's are Lie algebras, whereas GR's is the Dirac algebroid; 
this involves structure functions rather than structure constants.

\m 

\n The Constraint Closure Problem generalizes Kucha\v{r} and Isham's \cite{K92, I93} purely quantum-level Functional Evolution Problem facet of the Problem of Time 
to both Classical and Finite Theories as well as Quantum Field Theories.

\m 

%
\n{\bf Aspect 2: Assignment of Observables} (Sec 3 and Article VIII)  {\it Observables} are objects that brackets-commute with constraints.
These are useful objects due to their physical content, by which we seek to express a theory (or at least its physically meaningful outputs) in terms of observables.

\m 

\n{\bf The Problem of Observables} (facet 2) occurs when there are impasses to finding a sufficient set of these to do Physics; 
this is a common occurrence in Gravitational Theory, or (partially) Background-Independent Theory more generally.

\m

\n{\bf Taking Function Spaces Thereover} (strategy 2) resolves Assignment of Observables.   
At the classical level, and working piecemeal, this is largely trivial, referring to taking a suitably smooth set of functions over the unreduced phase space.  
However, on the one hand, Article IV explains how this becomes nontrivial at the quantum level.
On the other hand, in the presence of Closure, the function space in question must be restricted to suitably Lie-brackets-commute with the first-class constraints algebraic structure. 

\m 

\n The space of all possible observables of a given theory is thus of interest; this is a function space and furthermore itself an algebraic structure:  
the {\it observables algebraic structure}.
Each constraints algebraic substructure moreover induces a distinct notion of observables, each forming its own obervables algebraic superstructure.  
These constraints algebraic substructures form a bounded lattice, with the observables algebraic superstructures constituting the dual bounded lattice. 
Finally, the defining Poisson (or Dirac) brackets zero commutation condition can be rearranged to give an explicit PDE system, 
which can be approached using the Flow Method following from Lie's work (and Lagrange's earlier Method of Characteristics).   
This Flow Method reformulation, and solution of examples thereof \cite{PE-1, PE-2-3, DO-1, VIII} is a first major step beyond what was covered in \cite{ABook}.  

\m 

\n Constraints algebraic structures and observables algebraic structures are further spaces playing an underlying role as regards the nature of Physical Law.  
Each leads to its own notion of Tensor Calculus, for which the current Article furthermore provides a coordinate-independent notation to keep these clearly 
distinct from each other, and from their space and configuration space counterparts.

\m 


\n{\bf Aspect A: Constructability of Spacetime from Space} (Sec 4 and Article IX).  

\m  

\n{\bf The Spacetime Reconstruction Problem} (facet A) -- usually phrased at the quantum level \cite{Battelle} so it is explained in Article IV -- 
is resolved by Spacetime Construction.
A further well-known question of Wheeler's (II.25) can moreover be resolved with a classical-level spacetime construction 
from the assumption of just spatial structure; this Series concentrates upon Spacetime Construction in this sense.  

\m 

\n Both spacetime and specifics of GR-as-Geometrodynamics can be recovered as follows. 

\m 

\n A.i) {\bf Feeding a Deformed Family of Actions into Dirac's Algorithm} (via the constraints these actions encode: strategy 3). 

\m 

\n A.ii) {\bf Rigidity} \cite{Higher-Lie} then causes GR to emerge (strategy 3's victory condition), with very few alternatives alongside.   

\m 

\n For vacuum GR, this {\it derives} the explicit DeWitt form of GR-as-Geometrodynamics' kinetic term, as one of very few roots of an algebraic equation \cite{RWR, AM13}. 
The other roots return a generalized Galilean Geometrostatics, 
                                     Strong Gravity \cite{I76}, 
									 and constant mean curvature slicing: already familiar from treating GR's constraints as an initial-value problem \cite{York73}.   
Adding accompanying minimally-coupled matter moreover gives Einstein's dilemma between universal local Lorentzian and Galilean relativities 
- finite or infinite fundamental propagation speed - 
{\it as roots of an algebraic equation from the Dirac Algorithm, rather than a matter requiring an Einstein to intuit}. 
This moreover comes with the universal Carollian Relativity accompanying Strong Gravity as a third root, 
indicating the complete trilemma of finite, infinite, or zero fundamental propagation speed. 

\m 

\n More generally, Feeding a Family of Deformed Generators into a Lie (Brackets Consistency) Algorithm 
already has some capacity to pick out few significant structures from large input families, without having to involve Poisson brackets, or constraints, or Dirac's groundbreaking work. 
As such, Construction is already a property of Lie's Mathematics rather than just of Dirac's more specialized case. 
The example we use to demonstrate this is deriving Conformal and Projective Geometry as the only two possible flat geometries with quadratic generators in dimension $\geq 3$, 
i.e.\ a new Foundations of Geometry result closely parallelling the above `Einstein Dilemma' and `GR-as-Geometrodynamics' rigidities.  
This example is a second major result beyond \cite{ABook} itself, indeed exemplifying a further Background Independence aspect as follows. 

\m 

\n{\bf Aspect 3: Constructability of Space from Less Structured Space Assumed} (Sec 5 and Article IX). 

\m 

\n Both of our major new results announced so far can moreover be interpreted as further Foundations of Geometry following in parallel to 
A Local Resolution of the Problem of Time. 
By this, our classical local resolution succeeds in having unexpected applications in another major branch of Science: 
pure Geometry as well as the nature of Physical Law. 
This revelation \cite{PE-1, DO-1, A-Brackets} moreover begged for a comparative analysis to find the underlying cause. 
In turn, this revealed Lie's Mathematics to be a sufficient driving force for both, 
giving a further strong reason to write a series of Articles emphasizing the role of Lie's Mathematics in our classical local resolution.
So our classical local resolution has given new Foundations of Geometry, which, in turn, as a simpler arena, enabled clarification that `the hand at play' is Lie's, 
thus enabling this Series' both sharper and simpler pedagogy.  

\m 

%
\n Since GR is also a theory with a meaningful and nontrivial notion of spacetime, it has more Background Independence aspects than Relational Particle Mechanics does. 
Indeed, the Einstein field equations of GR determine the form of GR spacetime, as opposed to SR Physics unfolding on a fixed background spacetime.  
From a dynamical perspective, GR's geometrodynamical evolution {\sl forms spacetime itself}, 
rather than being a theory of the evolution of other fields {\sl on} spacetime or {\sl on} a sequence of fixed background spatial geometries.  
Regardless of whether spacetime is primary or emergent, there is now also need for the following.

\m 

%
\n{\bf Aspect 0$^{\prime}$: Spacetime Relationalism} (Sec 6 and Article X) by which the {\it diffeomorphisms} of spacetime itself, $Diff(\Frm)$, are physically redundant transformations.  

\m

\n{\bf Classical spacetime's invariance under spacetime diffeomorphisms} (strategy 0$^{\prime}$) straightforwardly implements this at the classical level. 

\m 

\n Quantum-level implementation is however harder, for instance feeding into the {\bf Measure Problem} of Path Integral Approaches to Quantum Gravity. 
Spacetime Relationalism is thereby indeed nontrivial.  

\m

%
\n{\bf Aspect 1$^{\prime}$: Spacetime Generator Closure} (Sec 7 and Article X) is classically implemented by 
{\bf infinitesimal spacetime diffeomorphisms closing as a Lie algebra} (strategy 1$^{\prime}$).  
This works out, at least at the classical level.  
At the quantum level, one may encounter a {\bf Generator Closure Problem} (facet 1$^{\prime}$).  

\m 

%
\n{\bf Aspect 2$^{\prime}$: Assignment of Spacetime Observables} (Sec 8 and Article X). 
If blocked, one has a {\bf Problem of Spacetime Observables}.
This is approached via {\bf Taking Function Spaces Thereover} (strategy 2$^{\prime}$), now over $PRiem(\Frm)$: 
the space of spacetime metrics over a fixed spacetime topological manifold $\Frm$.  

\m 

\n{\bf Aspect 3$^{\prime}$: Constructability of Spacetime from Less Structured Spacetime Assumed} (Sec 9 and Article X).  
This is novel to the current Article, the indefinite version of conformal versus projective flat geometry sufficing to exemplify it.  

\m 

%
\n N.B.\ that aspects 0$^{\prime}$ to 3$^{\prime}$ are absent from Isham and Kucha\v{r}'s {\it canonical-primality} classification.
These are now more finely and shaply split up than in \cite{APoT2, APoT3}, with \cite{ABook} intermediary in this regard. 
This is to parallel 1-and-2) to 3) and 4)'s split, space-time distinction causing there to be two separate Relationalisms 
in this case to spacetime having just the one.)

\m 

%
\n{\bf Aspect B: Foliation Independence} (Sec 8 and Article XII) Foliations of spacetime play major roles, both in dynamical and canonical formulations, 
and as a means of modelling the different possible fleets of observers within approaches in which spacetime is primary.  

\m 

\n The {\bf Foliation Dependence Problem} (facet B) is encountered if Foliation Independence cannot be established.    

\m 

\n Algebraically establishing the {\bf Refoliation Invariance} is Strategy B's victory condition for establishing this.  

\m 

\n N.B.\ Among our Constructability aspects, 3) and 3$^{\prime}$ each reside within a given primality -- space and spacetime respectively -- 
whereas A) is primality-traversing: a Wheelerian route from spatial to spacetime primality. 
B) is moreover also primality-traversing: the corresponding reverse Wheelerian route. 

\m 

%
\n All in all, the Problem of Time is a multi-faceted subset of the reasons why forming `Quantum Gravity' Paradigms is difficult and ambiguous; 
Further reasons are purely technical, or a mixture of both.

\m 

\n As we proceed along the current Article, each of the above implementations is moreover grounded in Lie's Mathematics in accord with this Series' claim. 
The generalizations required to extend these resolutions to solve combined rather than just piecemeal facets form the main subject of Articles VII to XIII. 

  
\section{Constraint Closure (Aspect 1)}\label{CC-Intro}

\subsection{Poisson brackets}\label{Poisson}

{\bf Definition 1} The {\it Poisson bracket}  
$\mbox{\bf \{} \m \mbox{\bf ,} \, \m  \mbox{\bf \}}$ of phase space functions $F(\biQ, \biP)$ and $G(\biQ, \biP)$ is given by
\be
\mbox{\bf \{} \, F \mbox{\bf ,} \, G \, \mbox{\bf \}} \:=  \frac{\pa F}{\pa \biQ} \cdot \frac{\pa G}{\pa \biP}  \m - \m  
                                                           \frac{\pa G}{\pa \biQ} \cdot \frac{\pa F}{\pa \biP}    \m . 
\label{PB}
\ee
{\bf Motivation} Poisson brackets are useful because of, firstly, turning out to afford a systematic treatment of constraints \cite{Dirac}. 

\m 

\n Secondly, they represent a preliminary step toward Quantization.    

\m 

\n{\bf Structure 1} In terms of Poisson brackets, the equations of motion are
\beq
\dot{\biQ} = \mbox{\bf \{} \, \biQ \mbox{\bf ,} \, H \, \mbox{\bf \}}                              \mma  
\dot{\biP} = \mbox{\bf \{} \, \biP \mbox{\bf ,} \, H \, \mbox{\bf \}}                              \m . 
\eeq
\n{\bf Remark 1} For any $F(\biQ, \biP, t)$, the total derivative 
$$ 
\frac{\d F}{\d t}  \es  \mbox{\bf \{} \, F \mbox{\bf ,} \, H \mbox{\bf  \}}  \m + \m  \frac{\pa F}{\pa t}  \m . 
$$ 
So if $F$ is not explicitly $t$-dependent, the intuitive conserved quantity condition 
\beq
0  \es  \frac{\d F}{\d t}  
\eeq 
becomes
\be              
\mbox{\bf \{} \, F \mbox{\bf ,} \, H \mbox{\bf  \}}  \es  0  \m .  
\eeq

\subsection{Preview of Constraint Closure's history as a Problem of Time facet}

The Constraint Closure Problem generalizes Kucha\v{r} and Isham's \cite{K92, I93} purely quantum-level Functional Evolution Problem facet of the Problem of Time 
to both Classical and Finite Theories as well as to QFTs.
Article IV explains what the Functional Evolution Problem is, and Articles IV and VI cover its generalization to Constraint Closure. 
For now, we go straight for the technically simpler classical manifestation of Constraint Closure.
Dirac envisaged this work \cite{Dirac51, Dirac58, Dirac}, whereas Article XIV points to Lie's \cite{Lie} partial precursor of it.

\subsection{Notions of equality}

\n{\bf Definition 1} Dirac's \cite{Dirac} notion of  
\be 
\mbox{\it weak equality } \m \approx  \m , 
\ee 
means equality up to additive functionals of the constraints.
In contrast, `strong equality' just means equality in the usual sense.  
The current Series also uses 
\be 
\mbox{`{\it portmanteau equality}' } \m \peq  
\ee 
to encompass both strong and weak equality; this is useful in more summarily introducing notions that have strong and weak versions.

\subsection{Two first instances of Dirac's multiplier-appending of constraints}

\n{\bf Remark 1} One part of Dirac's approach \cite{Dirac} to handling constraints is by appending them additively with Lagrange 
multipliers to a system's incipient or `bare' Hamiltonian, $H$ (see Sec \ref{Dir-Bra} for the other part).  

\m 

\n{\bf Definition 1} {\it Dirac's `starred' Hamiltonian} \cite{Dirac} is the result of appending a formalism of a theory's primary constraints $\bscP$ 
with a priori arbitrary phase space functions $A(\biQ, \biP)$ in the role of multipliers, 
\be 
H_{\sbiA\btcP}  \:=  H^*  
                \:=  H + \uc{\biA} \cdot \uc{\bscP}  \m . 
\ee 
$H^*$ is Dirac's historic notation, whereas $H_{\sbiA\btcP}$ reflects this notion's true-name {\it arbitrary-primary Hamiltonian}, 
signifying `with arbitrary multipliers appending primary constraints'.  
The square underbracket is the coordinate-free notation adopted for objects with indices running over some type of constraint, 
much as plain underline is for objects with spatial indices.  

\m 

\n{\bf Remark 2} Aside from such true names being much clearer and more deducible and memorable, 
in the current instance this frees Dirac's notation from `double booking' of star superscripts (because Dirac bracket is denoted by a star).  

\m 

\n{\bf Definition 2} {\it Dirac's total Hamiltonian} \cite{Dirac} is the result of appending these with {\it unknown functions} $\biu(\biP, \biQ)$ 
\be 
H_{\sbiu\btcP}  \:=  H_{\sT}  
                \:=  H + \uc{\biu} \cdot \uc{\bscP}      \m .
\ee 
Here again we are rewriting an historical notation of Dirac's, $H_{\sT}$, 
by a notation $H_{\sbiu\btcP}$ reflecting the true-name -- {\it unknown-primary Hamiltonian} -- 
signifying `with unknown multipliers appending primary constraints'.  

\m 

\n{\bf Remark 3} As an indication of how this is used (\cite{Dirac} in coordinate-free notation), 
\beq
0 \m \approx \m  \dot{\bscP}    
        \es      \mbox{\bf \{} \, \bscP \mbox{\bf ,} \, H_{\sbiu\btcP} \, \mbox{\bf \}}
        \es      \mbox{\bf \{} \, \bscP \mbox{\bf ,} \, H              \, \mbox{\bf \}}                  \m + \m  
		         \mbox{\bf \{} \, \bscP \mbox{\bf ,} \, \uc{\bscP}     \, \mbox{\bf \}} \cdot \uc{\biu}    \m .
\label{de-for-u}
\eeq
The first step is dictated by consistency, the second by Hamilton's equations, and the third by definition and linearity.
The brackets in use are, for now, Poisson brackets on unreduced phase space.  
The overall consequence is an explicit equation, with $\biu$ in role of unknowns, of course.

\subsection{Dirac's Little Algorithm}

\n{\bf Definition 1} {\it Dirac's Little Algorithm} (based Chapter 1 of \cite{Dirac}) 
consists of evaluating Poisson brackets between a given input set of constraints so as to determine whether these are consistent and complete.  
Four possible types of outcome are allowed in this setting.    

\m 

\n{\bf Type 0)} {\bf Inconsistencies}.
That this is possible in the Principles of Dynamics is clear from e.g.\ the Lagrangian 
\be 
L = \dot{q} + q                     \m ,
\ee 
giving as its Euler--Lagrange equations  
\be 
0 = 1                               \m . 
\ee
Dirac's envisaging the need to allow for the possibility of inconsistency allows for an algorithm with {\sl selection principle} properties.  

\m 

\n{\bf Type 1)} {\bf Mere identities} -- equations that reduce to 
\be 
0 \peq 0                       \m ,
\ee 
comprising strong identity 
\be 
0 = 0
\ee 
and the more general weak identity 
\be 
0 \approx 0                             \m.  
\ee 
\n{\bf Type 2)} {\bf Further secondary constraints}, i.e.\  equations independent of the Lagrange multiplier unknowns. 
Arising in this (implicitly variational) manner lies within the remit of secondary, rather than primary, constraints.

\m 

\n{\bf Type 3)} Relations amongst some of the appending Lagrange multipliers functions themselves. 
These are not constraints but a further type of equation, termed `{\bf specifier equations}' since they specify restrictions on the Lagrange multipliers; 
see Articles VII and IX for more on these.

\subsection{Discussion}

\n{\bf Remark 1} If type 0) occurs, the candidate theory is inconsistent. 

\m 

\n{\bf Definition 1} Let us refer to equations of all types arising from Dirac Algorithms bar 0) as {\it ab initio consistent}.

\m 

\n{\bf Definition 2} With type 1)'s mere identities having no new content, let us call types 2) to 4) `{\it nontrivial ab initio consistent objects}'.  
Note that we say `objects', not `constraints', to include type 3)'s specifier equations.  

\m 

\n{\bf Remark 2} If type 2) occurs, the resultant constraints are fed into the next iteration of Dirac's Algorithm, 
which starts with the extended set of objects.  
One is to proceed thus recursively until one of the following termination conditions is attained. 

\m  

\n{\bf Termination Condition 0) Immediate inconsistency} due to at least one inconsistent equation arising.

\m 

\n{\bf Termination Condition 1) Combinatorially critical cascade}. 
This is due to the iterations of the Dirac Algorithm producing a cascade of new objects 
down to the `point on the surface of the bottom pool' that leaves the candidate with no degrees of freedom 
(see the next version of Article XIV for depiction of pools and cascades).   
I.e.\ a combinatorial triviality condition.    

\m 

\n{\bf Termination Condition 2) Sufficient cascade}, i.e.\ running `past the surface of the bottom pool' of no degrees of freedom 
into the `depths of inconsistency underneath'.

\m 

\n{\bf Termination Condition 3) Completion} is that the latest iteration of the Dirac Algorithm has produced no new nontrivial consistent equations, 
indicating that all of these have been found. 

\m 

\n{\bf Remark 3} Our input candidate set of generators is either itself {\it complete} 
                                                     or {\it incomplete} -- `nontrivially Dirac' -- 
													 depending on whether it does not or does imply any further nontrivial objects.
If it is incomplete, it may happen that Dirac's Algorithm provides a completion, by only an {\it combinatorially insufficient cascade} arising, 
from the point of view of killing off the candidate theory.  
														 
\m 														 

\n{\bf Remark 4} So, on the left point of the trident, Termination Condition 3) is a matter of acceptance 
of an initial candidate set of constraints alongside the cascade of further objects emanating from it by Dirac's Algorithm.   
(This acceptance is the point of view of consistent closure; further selection criteria might apply.)    			
This amounts to succeeding in finding -- and demonstrating -- a `Dirac completion' of the incipient candidate set of constraints.  											 

\m  

\n{\bf Remark 5} On the right point of the trident, Termination Conditions 0) and 2) are a matter of rejection of an initial candidate set of constraints.  
The possibility of either of these applying at some subsequent iteration justifies our opening conception in terms of ab initio consistency.

\m 

\n{\bf Remark 6} On the final middle point of the trident, Termination Condition 1) is the critical case on the edge between acceptance and rejection; 
further modelling details may be needed to adjudicate this case.

\m 

\n{\bf Remark 7} In each case, {\it functional independence} \cite{Lie} is factored into the count made; 
the qualifiaction `combinatorial' indicates Combinatorics not always sufficing in having a final say.  
For instance, Field Theory with no local degrees of freedom can still possess nontrivial global degrees of freedom. 
Or Relationalism can shift the actual critical count upward from zero, e.g.\ by requiring a minimum of two degrees of freedom 
so that one can be considered as a function of the other.  
If this is in play, we use the adjective `relational' in place of (or alongside) `combinatorial'.  

\m 

\n{\bf Remark 8} If type 2)'s further constraints occur, these are fed into the subsequent iteration of the algorithm. 

\m 

\n This is by, firstly, defining $\bscQ$ as one's initial $\bscP$ 
alongside the subset $\bscR$ of the candidate theory's formulation's $\bscS$ that have been discovered so far, indexed by $\fQ = \fP \,\coprod \, \fR$.

\m

\n Secondly, by restarting from a more general form for our problem (\ref{de-for-u}) 
\beq
0  \m \approx \m  \dot{\bscQ}        
        \es       \mbox{\bf \{} \, \bscQ \mbox{\bf ,} \,  H_{\biu\btcQ} \, \mbox{\bf \}} 
        \es       \mbox{\bf \{} \, \bscQ \mbox{\bf ,} \,  H             \, \mbox{\bf \}}                   \m + \m  
		          \mbox{\bf \{} \, \bscQ \mbox{\bf ,} \,  \uc{\bscQ}    \, \mbox{\bf \}} \cdot \uc{\biu} 
   \m \approx \m  0                                                                                                                                             \m , 
\label{de-for-u-2}
\eeq
in what is hopefully by now self-explanatory notation.  

\m 

\n In detailed considerations, clarity is often improved by labelling each iteration's $\bscR$ and $\bscQ$ by the number of that iteration.
In the case of completion being attained, (the final $\bscR$) = $\bscS$ itself, whereas $\bscQ = \bscF$ 
(For now under the assumption explained below that all constraints involved are first-class: $\bscF$.)

\subsection{Each iteration's problem is a linear system}

\n{\bf Remark 1} (\ref{de-for-u}) or its subsequent-iteration generalization (\ref{de-for-u-2}) is a linear problem.  

\m 

\n Its general solution thus splits according to 
\be 
\biu  =  \bip + \biC                                                 \m , 
\ee
for particular solution $\bip$ and complementary function $\biC$.

\m 

\n By definition, $\biC$ solves the corresponding homogeneous equation  
\beq
\uc{\biC} \cdot \mbox{\bf \{} \, \uc{\bscC}  \mbox{\bf ,} \,  \bscP \, \mbox{\bf \}}  \m \approx \m  0   \m ,  
\eeq
where underbracket denotes constraint vector.

\m 

\n Furthermore, $\biC$ has the structure 
\be 
\uc{\biC}  \es  \uc{\bic} \, \cdot \, \uc{\uc{\biR}}_{\sS}                                                \m . 
\label{cR}
\ee 
The $\bic$ here are the totally arbitrary coefficients of the independent solutions. 
Also, $\biR$ is a mixed-index (and thus in general rectangular rather than square) matrix, 
whose second index runs over primary constraints while its first index runs over the generally-distinct independent solutions (hence the subscript S)  
Our general solution is next to be substituted into the total Hamiltonian, updating it.

\subsection{First- and second-class constraints}

\n{\bf Definition 1} {\it First-class constraints} \cite{Dirac, HTBook} 
\be 
\bscF \m \mbox{ indexed by } \m \fF
\ee  
are those that close among themselves under Poisson brackets.  

\m 

\n{\bf Definition 2} {\it Second-class constraints} \cite{Dirac, HTBook} 
\be 
\Sec \m \mbox{ indexed by } \m  \fE
\ee 
are defined by exclusion to be those that are not first-class. 

\m 

\n{\bf Remark 1} Our choice of notation is subject to the letter $\scS$ already being in use for secondary constraints.    

\m 

\n{\bf Diagnostic} For the purpose of counting degrees of freedom, 
first-class constraints use up two each whereas second-class constraints use up only one \cite{HTBook}.

\m 

\n{\bf Remark 3} First-class constraints are not necessarily gauge constraints;  
for now we give the canonical example of Dirac's Conjecture \cite{Dirac} failing \cite{HTBook, VII} as a counterexample.

\subsection{2 further instances of Dirac's multiplier-appending of constraints}

\n{\bf Definition 1} For later use, {\it Dirac's primed Hamiltonian} is 
\be 
H_{\sbip\btcP}  \:=  H^{\prime}  
                \:=  H           \m + \m  \uc{\bip} \cdot \uc{\bscP}                    \m ,
\ee  
for which we provide the true name {\it particular-primary Hamiltonian}, signifying `with particular-solution multipliers appending primary constraints'.  

\m 

\n{\bf Definition 2} Finally, {\it Dirac's extended Hamiltonian} is 
\be 
H_{\sE}  \:=  H  \m + \m  \uc{\biu} \cdot \uc{\bscP}  \m + \m  \uc{\bia} \cdot \uc{\bscS}     \m , 
\ee 
for arbitrary functions $\bia$ and specifically first-class secondary constraints $\bscS$.

\m 

\n{\bf Remark 1} Let us also reassign notation for this, now using $H_{\scF}$ standing for `Hamiltonian with first-class constraints appended', 
leaving it implicit that what multipliers can be solved for are.  

\m 

\n{\bf Remark 2} Such a notion could clearly be declared for each iteration of the Dirac Algorithm, 
with the above one coinciding with Dirac's Little Algorithm attaining completeness. 
In this sense, $H_{\scF}$ is a candidate theory's Hamiltonian that is {\it maximally} extended by appending of first-class constraints.   

\m 

\n{\bf Remark 3} While the {\sl name} `total Hamiltonian' is widely used in the literature, 
it is usually used loosely to mean various kinds of Hamiltonians including the `extended Hamiltonian'.
When examined in detail, moreover, one finds that it is the `extended Hamiltonian' notion \cite{HTBook} that enters GR. 
and, modulo Article VII's TRi adjustment, A Local Resolution of the Problem of Time.

\subsection{Removing second-class constraints}\label{Dir-Bra}

\n We are now to envisage the possibility of second-class constraints arising at some iteration in the Dirac Algorithm.

\m 

\n {\bf Motivation} Removal of second-class constraints 
is especially relevant since many standard quantum procedures are based on just first-class constraints remaining by that stage.
This usually entails classical removal of any other nontrivial consistent entities which feature in the original formulation.

\m 

\n{\bf Remark 1} Second-class constraints can moreover be slippery to pin down. 
This is because second-classness is not invariant under taking linear combinations of constraints. 
Linear Algebra dictates the invariant concept to be, rather, {\sl irreducibly second-class constraints} \cite{Dirac, HTBook} 
\be 
\bscI \m \mbox{ indexed by } \m \fI  \m .
\ee  
\n{\bf Proposition 1 (Dirac)} \cite{Dirac}. Irreducibly second-class constraints can be factored in by redefining the incipent Poisson bracket 
                                                                                                              with the  {\it Dirac bracket} 
\beq
\mbox{\bf \{} \,    F    \mbox{\bf ,} \, G    \, \mbox{\bf \}} \mbox{}^{\mbox{\bf *}}  \:=  \mbox{\bf \{} \,    F    \mbox{\bf ,} \, G    \, \mbox{\bf \}} \m - \m  
\mbox{\bf \{} \,    F    \mbox{\bf ,} \, \uc{\bscI}    \, \mbox{\bf \}}\cdot  \mbox{\bf \{} \,     \uc{\bscI}    \mbox{\bf ,} \, \uc{\bscI}    \mbox{\bf \}}^{-1}   \cdot
\mbox{\bf \{} \,    \uc{\bscI}    \mbox{\bf ,} \, G    \, \mbox{\bf \}}\m . 
\eeq
Here the --1 denotes the inverse of the given matrix, and each $\cdot$ contracts the underlined objects immediately adjacent to it.   

\m 

\n{\bf Remark 2} The role of classical brackets role initially played by the Poisson brackets may thus be taken over by the Dirac brackets.

\m 

\n{\bf Remark 3} Dirac brackets can moreover be viewed geometrically \cite{Sni} as more reduced spaces' incarnations of Poisson brackets.

\subsection{Dirac's Full Algorithm}

Proceed as before, except that whenever second-class constraints appear, one switches to (new) Dirac brackets that factor these in.  
This amounts to a fifth type of equation being possible, as follows.  

\m 

\n{\bf Type 4)} {\bf Further second-classness} can arise.

\m 

\n On the one hand, this could be {\it self-second-classness}, by which brackets between some new constraints do not close. 

\m 

\n On the other hand, it could be {\it mutually second-classness}, 
meaning that some bracket between a new constraint and a previously found constraint does not close. 
By which, this previously found constraint was just {\it hitherto first-class}.
I.e.\ first-classness of a given constraint can be lost whenever a new constraint is discovered.  

\m 

\n{\bf Remark 1} At each iteration, then, one ends up with a bare Hamiltonian with first-class constraints appended using multipliers.  
The final such is once again denoted by $H_{\scF}$, corresponding to having factored in all second-class constraints and appended all first-class constraints.  
Each other notion of Hamiltonian above can also be redefined for Dirac brackets, whether maximal or at any intermediary stage.   

\m 

\n{\bf Remark 2} This is as far as Dirac gets; subsequent discoveries in practise dicate the addition of the below sixth type. 
Dirac knew about this \cite{Dirac}, commenting on needing to be lucky to avoid this at the quantum level. 
But no counterpart of it enters his classical-level Algorithm.  

\m 

\n{\bf Type 5}) {\bf Discovery of a topological obstruction.} 
The most obvious examples of this are anomalies at the quantum level; it is however a general brackets phenomenon rather than specifically a quantum phenomenon.  

\m 

\n Two distinct strategies for dealing with this are as follows. 

\m 

\n{\bf Strategy i} Set a cofactor of the topological term to zero when the modelling is permissible of this. 
In particular {\bf strongly vanishing} cofactors allow for this at the cost (or discovery) of fixing some of the theory's hitherto free parameters.

\m 

\n{\bf Strategy ii} Abandon ship. 

\m 

\n{\bf Remark 3} In Dirac's Little Algorithm, everything stated was under the aegis of all objects involved at any stage are first-class, 
                                                                                         and that no topological obstruction terms occur.

\subsection{Constraint algebraic structures introduced}\label{NoC}

{\bf Structure 1} The end product of a successful candidate theory's passage through the Dirac Algorithm 
is a {\it constraint algebraic structure} consisting solely of first-class constraints closing under Poisson (or more generally Dirac) brackets. 

\m 

\n{\bf Modelling assumption} Assume that neither the topological nor tertiary complications explained below occur.

\m 

\n{\bf Structure 2} Under this condition, schematically, 
\beq
\mbox{\bf \{} \, {\bscF} \mbox{\bf ,} \,  {\bscF} \, \mbox{\bf \}} \speq  0                                                                     \m .
\label{F-F}
\eeq
This is a portmanteau for the strong version
\beq
\mbox{\bf \{} \, {\bscF} \mbox{\bf ,} \,  {\bscF} \, \mbox{\bf \}}   =  0                                                                       \m , 
\label{F-F-S}
\eeq
and the weak version: 
\beq
\mbox{\bf \{} \, \uc{\bscF}\mbox{\bf ,} \,  \uc{\bscF} \, \mbox{\bf \}} \es  \uc{\uc{\uc{\biF}}} \cdot \uc{\bscF}                                          \m .
\label{F-F-W}
\eeq
The $\biF$ here can be the structure constants of a Lie algebra, or a Lie algebroid's phase space functions,  
\be 
\biF(\biQ, \biP) \m . 
\ee

\subsection{Constraint Closure itself}\label{CC}

Given a set of constraints, are they a fortiori a complete and consistent set?  

\m 

\n On the one hand, suppose the initial set of constraints is of common or unstated provenance. 
Then Constraint Closure furthermore concerns whether the previous subsection's one-piece algebraic structure is realized.

\m

\n On the other hand, suppose the initial set of constraints come from two or more Constraint Providers. 
Then Constraint Closure concerns whether these sources are compatibly consistent and complete.  

\m 

\n More specifically, in the Background Independence and Problem of Time context, 
Constraint Closure is itself a {\it necessary test for candidate} Temporally and Configurationally Relational theories.
Given that we have these two constraint providers, our Dirac-type Algorithm splits, at least ab initio, 
into three separate checks. 

\m 

\n {\bf 1) Configurational Relationalism self-consistency} Whether our candidate $\bShuffle$ classical brackets self-closes.  
\be 
\mbox{\bf \{} \,  \bShuffle  \mbox{\bf ,} \,  \bShuffle \, \mbox{\bf \}} \speq  0                                                             \m ,   
\ee 
i.e.\ 
\be 
\mbox{\bf \{} \,  \bShuffle  \mbox{\bf ,} \, \bShuffle \, \mbox{\bf \}}  \es  0  
\ee 
in the strong case, or 
\be 
\mbox{\bf \{} \,  \uc{\bShuffle}  \mbox{\bf ,} \, \uc{\bShuffle} \, \mbox{\bf \}} \es  \uc{\uc{\uc{\biS}}} \cdot \uc{\bShuffle}  \m + \m  \m \uc{\uc{\biT}} \, {\Chronos}      \m .  
\label{Shuffle-Self}
\ee 
in the weak case, for structure functions $\biS$, $\biT$. 

\m 

\n{\bf 2) Mutual consistency between Configurational and Temporal Relationalisms} Whether $\bFlin$ and $\Chronos$ mutually close: 
the cross-brackets with one $\bFlin$ and one $\Chronos$ entry. 
\beq
\mbox{\bf \{} \, \bShuffle \mbox{\bf ,} \, \Chronos \, \mbox{\bf \}} \peqs   0 \mma  \Chronos \mbox{ is established as a good \m  $\lFrg$-object} \m .
\eeq
I.e. 
\beq
\mbox{\bf \{} \, \bShuffle \mbox{\bf ,} \, \Chronos \, \mbox{\bf \}}\m  \es  0  \m .
\eeq
in the strong case, or 
\beq
\mbox{\bf \{} \, \uc{\bShuffle} \mbox{\bf ,} \, \Chronos \, \mbox{\bf \}} \es  \uc{\uc{\biU}} \cdot \uc{\bShuffle}  \m + \m  \uc{\biV} \Chronos  \m .
\label{Mutual}
\eeq
in the weak case, for structure functions $\biU$, $\biV$. 

\m 

\n{\bf 3) Temporal Relationalism self-consistency} Whether $\Chronos$ self-closes. 
\beq
\mbox{\bf \{} \, \Chronos \mbox{\bf ,} \, \Chronos \, \mbox{\bf \}} \peqs  0  \m .  
\eeq
I.e.\  
\beq
\mbox{\bf \{} \, \Chronos \mbox{\bf ,} \, \Chronos \, \mbox{\bf \}}  \es   0  \m .
\eeq
in the strong case, or 
\beq
\mbox{\bf \{} \, \Chronos \mbox{\bf ,} \, \Chronos \, \mbox{\bf \}}  \es  \uc{\biW} \cdot \uc{\bShuffle}  \m + \m  X \, \Chronos  \m .
\label{Chronos-Self}
\eeq
in the weak case, for structure functions $\biW$, $X$. 

\m

\n{\bf Remark 1} The previous subsection's modelling assumption implies that $\bShuffle$ and $\Chronos$ are both first-class, $\bscF$.   
In particular, this means that 
\be 
\bShuffle  \es  \bFlin \m . 
\ee 
\n{\bf Remark 2} In common examples this can be concatenated with 
\be 
\bFlin \m \Rightarrow \m \bGauge \m 
\ee 
-- an equality known as {\it Dirac's Conjecture} \cite{Dirac} -- by which our initial candidate $\bShuffle$ would have gauge constraint status, 
Dirac's Conjecture is, however, in general false (see \cite{HTBook, ABook} and Article VII): 
\be 
\bFlin \m \not{\Rightarrow} \m \bGauge \m 
\ee 
so $\bShuffle$ will not in all cases constitute gauge constraints $\bGauge$.  

\m 

\n{\bf Remark 2} Once one considers whichever combination of GR, Background Independence, or whole-universe theories, 
what is meant by `gauge' picks up subtleties beyond what occurs in flat spacetime's Electromagnetism or Yang--Mills Theory.
This is to the extent that what is `gauge' in GR remains, even to date, an unsettled matter (see Article VII for further details). 

\m 

\n{\bf Remark 3} Even if one steers clear of any of the above three factors, moreover, Dirac's Conjecture remains in general false. 
Yet awareness of this, and its consequences, does not look to be widespread in the current Particle Physics community, 
or even in most parts of the current wider Theoretical Physics community. 

\m 

\n{\bf Restriction 1} $\biT = 0$ means that (\ref{Shuffle-Self}) self-closes in a manner leaving Configurational Relationalism 
pairwise compatible with Constraint Closure.  
Configurational Relationalism here realizes the $\lFrg$ group.  

\m 

\n{\bf Restriction 2} $\biW = 0$ means that $\Chronos$ self-closes in a manner leaving Temporal Relationalism pairwise compatible with Constraint Closure.
Temporal Relationalism here realizes some group  
\be 
\lFrg_{\sT}  \m ; 
\ee 
in the finite case, this moreover has just one generator, and so must be an Abelian group.  

\m 

\n{\bf Restriction 3} Suppose Restriction 1 holds, and also $\biU = 0$.  
Then $\Chronos$ is a good $\lFrg$ object. 

\m 

\n{\bf Restriction 4} Suppose Restriction 2 holds, and also $\biV = 0$.  
Then $\bShuffle$ is a good $\lFrg_{\sT}$ object. 

\m 

\n{\bf Restriction 5} Suppose all of the previous 4 restrictions hold: 
\b 
\biT    = 
\biU    = 
\biV    = 
\biW    = 0  \m .   
\ee 
Then Temporal and Configurational Relationalism are totally decoupled from each other. 

\m 

\n{\bf Structure 1} At the level of the constraint algebraic structure, this is reflected by the {\it direct product form}:
\be 
\FrF  \es  \lFrg_{\sT} \times \lFrg   \m . 
\label{dir}
\ee 
{\bf Structure 2} Restriction 3 amounts to {\it semidirect product form} in one direction, 
\be 
\FrF  \es  \lFrg_{\sT} \rtimes \lFrg  \m , 
\ee 
and Restriction 4 to the same in the other direction 
\be 
\FrF  \es  \lFrg_{\sT} \ltimes \lFrg  \m .  
\ee 
\n{\bf Remark 4} $\biT \neq 0$ would leave Temporal        Relationalism as an implied as a {\it self-integrability} of Configurational Relationalism, 
       whereas   $\biW \neq 0$ would leave Configurational Relationalism as an implied as a self-integrability of Temporal        Relationalism. 

\m 	    

\n{\bf Structure 3} We denote the above two {\it one-way integrabilities} by, respectively,  
\be 
\FrF  \es  \lFrg_{\sT} \Thomas \lFrg    
\ee
and 
\be 
\FrF  \es  \lFrg_{\sT} \LThomas \lFrg  \m .  
\ee
\n{\bf Example 1} As Article VII details, Thomas precession amounts to a well-known prototype of Structure 3.

\m
 
\n{\bf Structure 4} Both of the above integrabilities occurring concurrently would amount to a {\it two-way integrability}: 
Temporal and Configurational Relationalisms concurrently being implied as integrabilities of each other.
We denote this by 
\be
\FrF  \es  \lFrg_{\sT} \TwoWay \lFrg   \m .  
\label{2-way}
\ee 
{\bf Remark 5} Tertiary constraints could moreover appear, in any combination of eqs (\ref{Shuffle-Self}, \ref{Mutual}, \ref{Chronos-Self}). 
This could amount to an at least ab initio failure of either Temporal or Configurational Relationalism's compatibilities with Constraint Closure, 
or of both concurrently: a mutual 3-facet failure. 
This is a matter of Configurational and Temporal Relationalism proving to be {\it incomplete} as regards direct provision of all of a theory's constraints.

\m 

\n Tertiary constraints would come with the extra problems of, on the one hand, whether their presence is to extend an originally postulated group.

\m 

\n On the other hand, of considering what happens with {\sl their} self-brackets and mutual brackets with each of the blocks of constraints 
from ab initio Configurational and Temporal Relationalism. 
These could of course produce further tertiary constraints of their own, and so on: a so-called `cascade' of constraints. 

\m 

\n At any stage, further second-class constraints could appear among the new tertiaries.
This can include rendering constraints from some previous iteration second-class relative to new tertiaries.
This is dealt with by reassigning Dirac brackets as many times as necessary; 
note that this is capable of shrinking an ab initio $\lFrg$, 
or of rendering Temporal Relationalism already directly encoded within the new Dirac brackets algebraic structure.

\m 

\n{\bf Remark 6} Such a cascade could be sufficient to induce triviality, or even inconsistency. 

\m 

\n{\bf Remark 7} Specifier equations could instead arise at whichever iteration of the Dirac Algorithm. 

\m 

\n{\bf Remark 8} Henneaux and Teitelboim emphasize and largely illustrate \cite{HTBook} 
how all combinations of first and second class constraints and primary and secondary constraints are possible in whichever steps of the Dirac Algorithm.  
A few examples in this regard can be found in Sec VII.4-6. 

\m

\n{\bf Remark 9} (\ref{dir}-\ref{2-way}) constitute a characterization of the extent to which an a priori split of a brackets algebra into 2 blocks 
is respected by the algebra's relations.  
This has various other well-known applications, such as Lie group contractions \cite{Gilmore} 
and the occurrence of decomposition into qualitatively differently different blocks in Lie Theory \cite{Gilmore, Serre-Lie, FHBook}.  

\m  

\n{\bf Remark 10} Articles II to IV moreover do not contain any examples realizing topological obstructions, cascades, or specifier equations.
This is with the sole exception of Spacetime Construction, though this case can however be avoided by suppressing terms by strong vanishing.  
As such, Remarks 5 to 10 can be viewed as just preliminary set-up for Article VII rather than essential to the current Article; 
less experienced readers should thus for now not worry about these Remarks.

\subsection{Constraint Closure Problem}

For Poisson brackets -- or, in presence of secondary constraints, the final Dirac brackets -- the following problems could occur. 

\m 

\n{\bf Constraint Closure Problem 1)} This arises if the generators fail to close due to unexpected brackets obstruction terms arising; 
while this is best known in the quantum-level occurrence of anomalies \cite{Dirac, AW84, Bertlmann}, classical counterparts are also possible.  
This is a type of inconsistency, and more specifically, a `hard obstuction' that is `topologically rooted'. 
These are successively superior monickers to `1', and so subsequent mention is of {\bf Constraint Closure Problem by Topological Obstruction Inconsistency}. 
Further topological considerations are postponed to Article XIV.  

\m 

\n{\bf Constraint Closure Problem 2)} This consists of tertiary constraints arising.  

\m 

\n This is not by itself fatal to Constraint Closure, but can reveal Closure-to-Relationalism incompatibilities. 
Or, at least, incompatibilities with any principles necessitating a particular group $\lFrg$, 
since tertiary constraints arising can necessitate extending this to some $\lFrg^{\prime}$.  

\m 

\n This particular example is an `enforced group extension' problem, 
which may become severe if the original group is held to arise from some inviolable higher principle.
As such, we subsequently refer to this as the {\bf Constraint Closure Problem by Enforced Group Extension}.  

\m

\n More generally, tertiary constraints arising signifies that our Relational Constraint Providers have failed to directly arrive at all constraints.  

\m 

\n Tertiary second-class constraints can moreover include rendering directly provided hitherto expectedly first-class constraints second-class. 
This gives, for example, another way for a given $\lFrg$ to fail: {\bf Constraint Closure Problem by Enforced Group Reduction}.  
Or, as a second example, a distinct way for our encodement of Temporal Relationalism failing.  
A further way of the latter occurring is if $\Chronos$ is, or through tertiary constraints, becomes a 1-way integrability of (perhaps a new, adjusted) $\lFrg$. 
In such a case, Configurational Relationalism may have the capacity to rob $\Chronos$ of coprimary status; 
interestingly, Supergravity can be interpreted in this manner \cite{AMech, ABook}.  

\m 

\n All in all, Problem 2)'s more general name is {\bf Constraint Closure Problem by Tertiary Incompleteness}.  

\m 

\n{\bf Constraint-and-Specifier Closure Problem} This consists of specifier equations arising from the Dirac Algorithm, 
by which the Constraints category of our ab initio Closure Problem is itself broken. 
This is an incompleteness problem.

\m 

\n{\bf Constraint Closure Problem 3)}  This arises if Problem 2) causes a sufficient cascade to leave us with no degrees of freedom, or inconsistency.   
This is a `death by 1000 cuts' type phenomenon, though the role of the 1000 is played, more precisely, by the number of iterations in the Dirac Algorithm. 
As such, `sufficient cascade' is a truer name, and we subsequently use {\bf Constraint Closure Problem by Sufficient-Cascade Inconsistency}. 

\m 

\n{\bf Constraint Closure Problem 4} This arises if two-way integrability occurs, either directly, or once tertiary blocks are in play. 
This amounts to {\bf Temporal and Configurational Relationalism ceasing to be separable notions due to Algebraic Interference}.

\subsection{Sideboard of Strategic Elements for dealing with Closure Problems}

\n{\bf Strategic Element 1)} {\bf Abandon} one's candidate theory. 

\m 

\n{\bf Strategic Element 2)} {\bf Avoid} topological obstructions, 
                                        specifiers, 
										sufficient cascades, 
										or even any tertiaries at all when required, 
in those case that all such are accompanied by factors that can {\bf strongly vanish}.  

\m 

\n{\bf Strategic Element 3)} Suppose terms in candidate Lagrangian source topological obstructions, 
                                                                         specifiers, 
                                                                         sufficient cascades, 
												   			   		  or even any tertiaries running contrary to one's principles. 
Then {\bf Avoidance by removing or adding terms} gives another alternative; adding can work by cancelling contributions. 
This works provided that we are left with {\sl some} terms in the Lagrangian, 
and that this does not run contrary to any `commensurate or higher' principles we require.  

\m     

\n{\bf Strategic Element 4)} {\bf Accept Incompatibility}: that Incompleteness can amount to a given $\lFrg$, or group action thereof, 
being rejected by a candidate theory's configuration space $\FrQ$.  

\m 

\n Failure of Configurational Relationalism self-consistency may moreover be acceptable 
if one were not to be operating with principled reasons to adhere to $\lFrg$ in the role of gauge group. 

\m 

\n This permits both enhanced and reduced groups in place of $\lFrg$. 

\m 

\n{\bf Strategic Element 5)} A reverse operation to Constraint Provision is the {\bf encoding of constraints}.  
I.e.\ upon finding constraints, one aims to subsequently build auxiliary variables into the theory's action. 

\m 

\n See Article VII for further details of these strategies, as well as futher variants.

\subsection{Finite theory simplification}

\n{\bf Lemma 1} Temporal Relationalism self-consistency is automatic for finite quadratic theories.

\m 

\n{\u{Proof}} These only have one single-component constraint. 
But any single component object strongly commutes with itself.  $\Box$ 

\m 

\n{\bf Remark 1} This enforces $\biW = 0$, $X = 0$ in this case.  

\m 

\n{\bf Remark 2} As we shall see in Section \ref{GR-Ex}, however, this argument breaks down upon passing to Field Theory.

\subsection{Lattices from Dirac's Algorithm}

\n{\bf Remark 1} We take lattices, and Order Theory \cite{Lattices} more generally, 
to be a standard occurrence in the theory of Lie algebras and Lie groups at least since Serre \cite{Serre-Lie}; see Article XIV for further details. 

\m 

\n{\bf Structure 1} For subsequent use in this Series, there is further interest in finding the (in particular conceptually meaningful) 
consistent subalgebras supported by the full constraint algebra. 
The notions of constraint in question form a bounded lattice, 
\be 
\lattice_{\btcC}
\ee 
with all first-class constraints $\bscF$ as top element, and the absence of constraints as bottom element.
The other members of the lattice of notions of constraint are middle elements, the {\it notions of Z-algebraic structure} denoted by 
\be 
\bscZ \m \mbox{ with each type indexed by } \m  \fZ  \m .
\ee  
See row 2 of Fig \ref{C-Latt} for a schematic sketch. 
Thus $\bscC$ comprises $\emptyset$, $\bscZ_{\sfZ}$ and $\bscF$, arranged to form $\lattice_{\btcC}$.

\subsection{RPM example}

{\bf Example 1} Euclidean RPM's constraint algebra's nonzero Poisson brackets are 
\beq 
\mbox{\bf \{} \,  \u{\sbcL} \mbox{\bf ,} \,  \u{\sbcL} \, \mbox{\bf \}} \es  \u{\u{\u{\epsilon}}} \cdot \u{\sbcL} \mma
\ee
\be 
\mbox{\bf \{} \,  \u{\bscP} \mbox{\bf ,} \,  \u{\sbcL} \, \mbox{\bf \}} \es  \u{\u{\u{\epsilon}}} \cdot \u{\bscP} \m ,
\label{ERPM-Cons}
\eeq
where $\epsilon$ is 3-$d$ space's alternating tensor.  

\m 

\n{\bf Remark 1} The first of these means that the $\u{\sbcL}$ close as a Lie algebra, 
which is a subalgebra of the full constraint algebra (itself a larger Lie algebra in this case).

\m 

\n{\bf Remark 2} The second signifies that $\u{\bscP}$ is a `good object' -- in this case a vector -- under the rotations generated by the $\u{\sbcL}$.

\m 

\n{\bf Remark 3} Together, the first and second equations correspond to no obstructions, tertiaries, or specifier equations arising, and $\biT = 0$.  
Self-consistency of $\lFrg = Eucl(d)$ is confirmed.

\m 

\n{\bf Remark 4} $\scE$ additionally closes with these gauge constraints, 
in a manner that establishes it as a scalar under the corresponding $Eucl(d)$ transformations.
This corresponds to no obstructions, tertiaries, or specifier equations, and $\biV = 0$.  

\m 

\n{\bf Remark 5} Lemma 1 finishes matters off, corresponding to no obstructions, tertiaries, or specifier equations, as well as $\biW = 0$, $X = 0$ holding.
 
\m 
 
\n{\bf Remark 6} Euclidean RPM thus avoids all parts of the Constraint Closure Problem.     
It exhibits moreover independence of Temporal and Configurational Relationalisms: either can be `switched off' without affecting the other. 
This accounts for our prior mentions of each of merely Temporally-Relational and merely Configurationally-Relational Particle Mechanics.

\subsection{GR Example}\label{GR-Ex}

{\bf Example 2} For full GR, the constraints close \cite{Dirac51, Dirac58, Tei73} in the form of the {\it Dirac algebroid} 
\be 
\FrD\mbox{irac}(\bupSigma) \m , 
\ee 
whose coordinate-independent form is 
\be
\mbox{\bf \{} \, ( \u{\bscM}  \,  |  \,  \u{\bupxi}  ) \mbox{\bf ,} \, (  \u{\bscM}  \,  |  \,  \u{\bupchi}  )  \, \mbox{\bf \}}  \es   (  \u{\bscM}    \,  |  \, \, \u{[ \bupxi, \bupchi ]}  )    \m ,
\label{Mom,Mom}
\ee
\be
\mbox{\bf \{} \, (  \scH    \,  |  \,  \upmu  ) \mbox{\bf ,} \, (  \u{\bscM}  \, | \,  \u{\bupxi}  ) \, \mbox{\bf \}}  \es   (  \pounds_{\u{\sbupxi}} \scH  \,  |  \,  \upmu  )  \m , 
\label{Ham,Mom}
\ee
\be 
\mbox{\bf \{} \, (  \scH    \,  |  \,  \upzeta  ) \mbox{\bf ,} \, (  \scH  \, | \,  \upomega  ) \, \mbox{\bf \}}  \es   
( \u{\bscM} \cdot \u{\u{\bh}}^{-1} \cdot \, | \, \upzeta \, \overleftrightarrow{\u{\bpa}} \upomega ) \m . 
\label{Ham,Ham}
\ee
$( \m | \m )$ is here the integral-over-space functional inner product, 
$[ \m , \m ]$, the differential-geometric commutator Lie bracket, 
and $\bupxi$, $\bupchi$, $\zeta$ and $\upomega$ are smearing functions. 
Such `multiplication by a test function' serves to render rigourous a wider range of 
     `distributional' manipulations \cite{AMP} provided that these occur under an integral sign.  

\m 

\n{\bf Remark 1} This closes in the sense that there are no further constraints or other conditions arising in the right hand side expressions. 
The Constraint Closure Problem is thus a {\sl solved} problem at the classical level for full GR.  

\m 

\n{\bf Remark 2} The first Poisson bracket means that $Diff(\bupSigma)$ on a given spatial hypersurface themselves close as an (infinite-$d$) Lie algebra.  
This corresponds once again to no obstructions, tertiaries, or specifier equations arising, and $\biT = 0$.    
 
\m 

\n{\bf Remark 3} The second signifies that $\scH$ is a good object -- a scalar density -- under $Diff(\bupSigma)$.  
This corresponds once again to no obstructions, tertiaries, or specifier equations arising, and $\biV = 0$.  
 
\m 

\n{\bf Remark 4} Both Remarks 2 and 3 are kinematical rather than dynamical.  
The third Poisson bracket is however both dynamical and more complicated in form and meaning \cite{Tei73}.
In particular, while no obstructions, tertiaries, or specifier equations arise here either, and $X = 0$, $\biW \neq 0$, 
by which $\u{\bscM}$ is an integrability of $\scH$ \cite{MT72}.  

\m

\n{\bf Consequence i)} So if one tried to consider $\Riem(\bupSigma)$ without $Diff(\bupSigma)$ being physically irrelevant, 
(\ref{Ham,Ham}) would in any case enforce this.
Thereby, neither GR's $\scH$, nor its underlying Temporal Relationalism, can be entertained without $\u{\bscM}$ 
or its underlying Configurational Relationalism (see Article IX for further details).   
This to be contrasted with how Temporal and Configurational Relationalism can be entertained piecemeal in RPM.  
The GR case clearly has a further triple facet interference to handle at this point.  

\m 

\n{\bf Remark 5} These are moreover not structure constants but structure functions, out of containing  $\bh^{-1}(\bh(\underline{x}))$.  

\m 

\n{\bf Consequence ii)} The transformation itself depends on the object acted upon \cite{Tei73}, in contrast with the familiar case of the rotations.  

\m 

\n{\bf Consequence iii)} The GR constraints form a more general algebraic structure than a Lie algebra: a {\it Lie algebroid}.   
More specifically, (\ref{Mom,Mom}--\ref{Ham,Ham}) form the Dirac algebroid \cite{Dirac51, Dirac58} $\FrD\mbox{irac}(\bupSigma)$.
In the Theoretical Physics iterature, Bojowald raised awareness of this subtlety \cite{BojoBook}.  
Because of this occurrence, we introduce the portmanteau {\sl Lie algebraic structure}, to jointly encompass Lie algebras and Lie algebroids; 
{\it constraint algebraic structures} are meant in this sense.  

\m 

\n{\bf Consequence iv)} By not forming a Lie algebra, the GR constraints clearly form a structure other than $Diff(\Frm)$.  
%
{\begin{figure}[!ht]
\centering 
\includegraphics[width=1\textwidth]{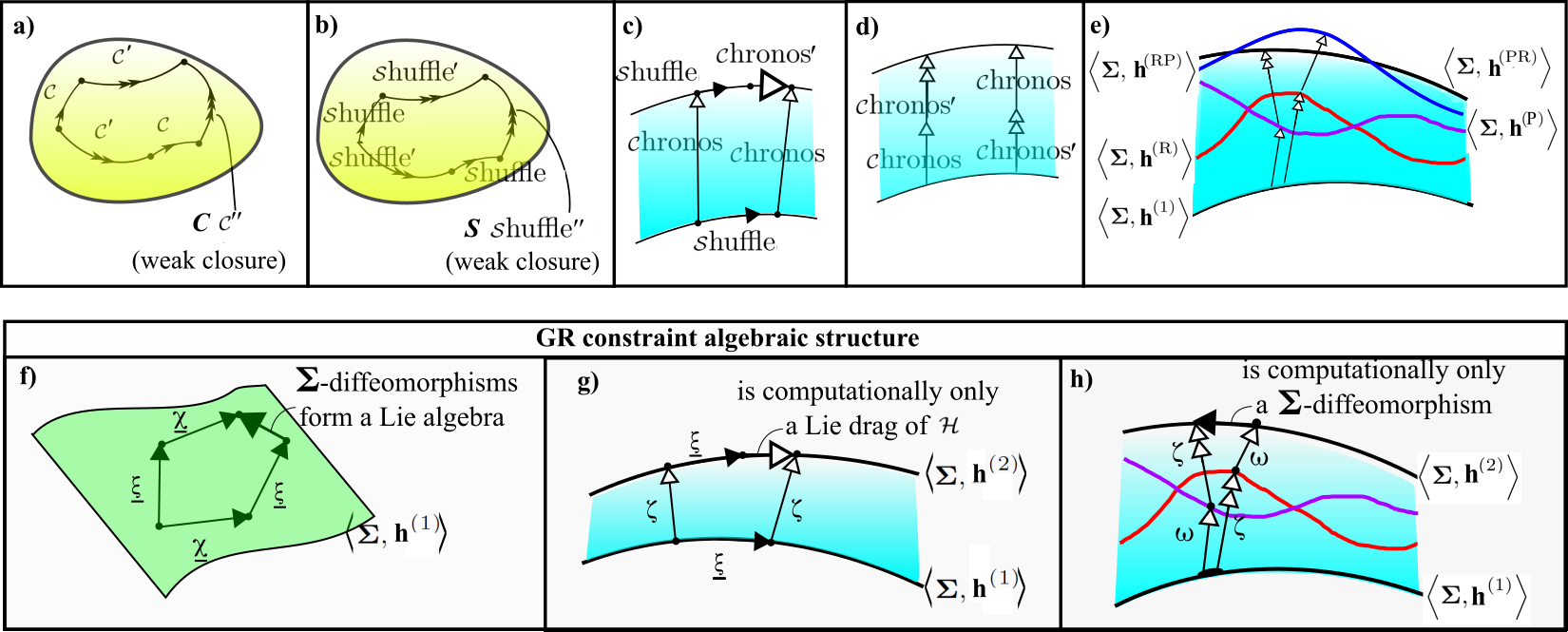}
\caption[Text der im Bilderverzeichnis auftaucht]{\footnotesize{a) and b) depict the algebraic nature of Constraint Closure 
for general first-class constraints and shuffle constraints respectively. 
[Algebra and group commutation relation diagrams are easy to pick out in this Series of Articles's presentation 
due to being drawn upon lime-green egg-shaped spaces.]  

\m 

\n c) $\Chronos$ as a good $Diff(\bupSigma)$ object: these commute up to a possible second use of $\Chronos$.  

\m 

\n d) In Finite Theories, the bracket of two $\Chronos$ constraints is strongly zero. 

\m 

\n e) In Field Theories, however, this bracket is in general nonzero.  

\m 

\n f) to h) then depict the specific case of GR's Constraint Closure. 
The meanings of Figs e) to h) are further explored in Sec \ref{Fol}.}} 
\label{CC-Figure}\end{figure}} 
 
\m 

\n{\bf Example 3)} Minisuperspace (spatially homogeneous GR) \cite{Magic} has a single finite constraint, $\scH_{\sm\si\sn\si}$.
It thus closes trivially under the Abelian constraint algebra,   
\beq
\mbox{\bf \{} \,     \scH_{\sm\si\sn\si}     \mbox{\bf ,} \, \scH_{\sm\si\sn\si}    \, \mbox{\bf \}}  \es  0 \m .
\eeq  
{\bf Remark 7} (\ref{Mom,Mom}--\ref{Ham,Ham}) has here simplified via the spatial covariant derivative $\bcalD$ now annihilating everything by homogeneity.

\m 

\n This causes, firstly, there to be no momentum constraint in the first place.  

\m 

\n Secondly, it gives another argument (to Lemma 1) for why the bracket of two Hamiltonian constraints vanishes.

\m

\n{\bf Remark 8} Minisuperspace thus clearly passes all kinds of Constraint Closure.

\m  

\n{\bf Remark 9} See Article VII for the further examples of Constraint Closure in Electromagnetism, Yang--Mills Theory, these coupled to GR, and Strong Gravity. 
For now, we summarize the current Article's examples in Fig \ref{C-Latt}.

\subsection{Constraint algebraic structures exemplified}

\n{\bf Structure 1}  The {\it full space of classical first-class constraints} is\footnote{Suppressing less notation, 
this is the constraint algebraic structure $\bFrF(\Phase(\lFrs), H)$ and likewise for subsequent constraint algebraic structures listed here.}
\be 
\bFrF                         \m ,
\ee 
The {\it space of absence of constraints} is just $id$.  
The {\it space of classical first-class linear constraints} is 
\be 
\bFrF\ml\mi\mn                \m ,
\ee 
the {\it space of classical gauge constraints} is 
\be 
\bFrG\ma\muu\mg\me                \m ,
\ee 
and the {\it space of Chronos constraints} is  
\be 
\bFrC\mh\mr\mo\mn\mo\ms       \m . 
\ee
{\bf Structure 2} The totality of constraints subalgebraic structures for a given formalism of a given theory  
\be
\mbox{bounded lattice } \m \lattice_{\sFrC}   \m . 
\ee
The identity algebraic structure is the bottom alias zero element, and 
the full algebraic structure of first-class constraints is the bottom alias zero element. 
All other elements are middle elements: the {\it Z constraint algebraic structures}, denoted by 
\be 
\bFrZ \m \mbox{ with each type indexed by } \m \fZ  \m . 
\ee
Thus $\bFrC$ comprises $id$, $\bFrZ_{\sfZ}$ and $\bFrF$, arranged to form $\lattice_{\sFrO}$.  

\m 

\n{\bf Remark 1} Constraint algebraic structures  are comparable to configuration spaces $\FrQ$ and phase spaces $\Phase$ in the study of the nature of Physical Law, 
and whose detailed structure is needed to understand any given theory.
This refers in particular to the topological, differential and higher-level geometric structures observables algebraic structures support, 
now with also function space and algebraic levels of structure relevant. 

\m 

\n{\bf Remark 2} This means we need to pay attention to the Tensor Calculus on constraint algebraic structures as well, 
justifying our use of underbrackets to keep constraint-vectors distinct from spatial ones.  

\m 

\n{\bf Remark 3} See Article VII for a specifically TRi Constraint Closure.  
%
%
%
{            \begin{figure}[!ht]
\centering
\includegraphics[width=0.7\textwidth]{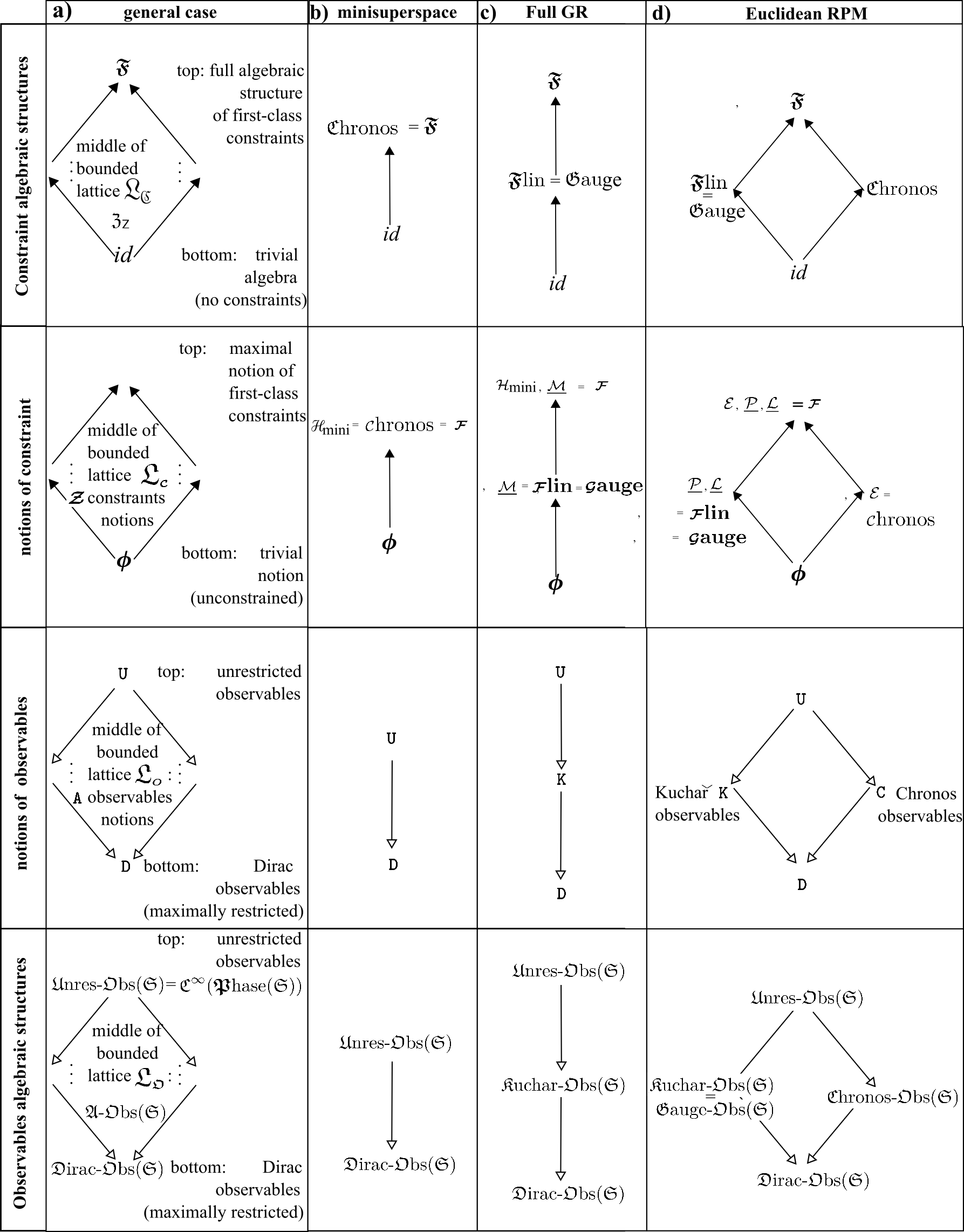}
\caption[Text der im Bilderverzeichnis auftaucht]{ \footnotesize{Lattices of notions                 of constraints (row 2)
                                                                      and of constraint subalgebraic structures     (row 1), 
												   with the dual lattices of notions                 of observables (row 3) 
                                                                      and of observables subalgebraic structures    (row 4).  
%
																	  
\m 
																	  
\n a) in general, schematically. 

\m 

\n b) For minisuperspace GR. 

\m 

\n c) For full GR: a first arena with a nontrivial middle. 

\m 

\n d) for Euclidean RPM: a first arena with a nontrivial-poset middle rather than just a chain.  } }
\label{C-Latt}\end{figure}            }

\vspace{10in}

\section{Assignment of Observables (Aspect 2)}\label{EitoO}

\subsection{Unrestricted observables}

\n{\bf Structure 1} The most primary notion of observables involves, given a state space $\FrS(\lFrs)$ for a system $\lFrs$, 
{\it Taking a Function Space Thereover},  $\FrF(\FrS(\lFrs))$. 
In the canonical setting, $\FrS = \Phase(\lFrs)$.  

\m 

\n{\bf Structure 2} Let us first consider this in the unrestricted case, meaning, in the canonical setting, in  the absense of constraints. 
The {\it unrestricted observables}  
\be
\sbiU(\biQ, \, \biP)                                        
\ee 
form, if working over the smooth functions, say, the {\it space of unrestricted observables}   
\be 
\UnresObs(\lFrs) \es \FrC^{\infty}(\Phase(\lFrs))  \m .
\ee 
this is clearly a function space.
At the classical level, this is a trivial extension of the theoretical framework; 
at the quantum level, however, self-adjointness and Kinematical Quantization already impinge at this stage.

\m 

\n{\bf Notation} In this Series, we use typewriter font to pick out notions of observables, 
and denote the corresponding spaces of observables by $\Obs$ with suitably descriptive prefixes.

\subsection{Constrained observables}\label{CA}

{\bf Structure 1} Facet interference with the presence of constraints $\bscC$ imposes the {\it commutant condition}  
\be 
\mbox{\bf \{} \,  \bscC  \mbox{\bf ,} \, \sbiO \, \mbox{\bf \}}  \speq  0  
\ee
on canonical observables functions, $\sbiO$. 
I.e.\
\be 
\mbox{\bf \{} \,  \bscC  \mbox{\bf ,} \, \sbiO \, \mbox{\bf \}}   \es   0  
\ee
in the strong case, or 
\be  
\mbox{\bf \{} \,  \uc{\bscC}  \mbox{\bf ,} \, \uo{\sbiO} \, \mbox{\bf \}}   \es  \m \uc{\uo{\uc{\biW}}} \cdot \uc{\sbiC}
\ee 
in the weak case, with structure constants $\biW$. 
The undertilde constitutes coordinate-free notation for an index running over some type of observables.  
 
\m 

\n{\bf Remark 1} This commutant condition is clearly a Lie brackets relation. 

\m 

\n{\bf Motivation 1} In a constrained theory, 
constrained observables are more physically useful than just any functions (or functionals) of $\biQ$ and $\biP$, 
due to their containing between more, and solely, physical information.    

\m 

\n{\bf Lemma 2} \cite{AObs} For this commutant condition to be consistent \cite{AObs}, 
the $\bscC$ must moreover be a closed subalgebraic structure of first-class constraints $\bscF$.  

\m 

\n{\u{Proof}} This follows from the Jacobi identity 
\be
  \mbox{\bf \{} \, \sbiO  \mbox{\bf ,}  \mbox{\bf \{} \, \bscC  \mbox{\bf ,} \, \bscC  \, \mbox{\bf \}}\, \mbox{\bf \}}     \es 
- \mbox{\bf \{} \, \bscC  \mbox{\bf ,}  \mbox{\bf \{} \, \bscC  \mbox{\bf ,} \, \sbiO  \, \mbox{\bf \}} 
- \mbox{\bf \{} \, \bscC  \mbox{\bf ,}  \mbox{\bf \{} \, \sbiO  \mbox{\bf ,} \, \bscC  \, \mbox{\bf \}}\, \mbox{\bf \}}    \speq  0     \m .  \m  \Box
\ee  
\n{\bf Remark 1} This further Lie-algebraic relation {\sl decouples} Assignment of Canonical Observables to occur {\sl after} 
establishing Constraint Closure of the constraints provided by 
Temporal and Configurational Relationalism, as depicted in Figure I.2.  

\m  

\n{\bf Structure 1} The opposite extreme to imposing no restrictions is to impose all of a modelling situation's first-class constraints.
This gives the {\it Dirac observables} $\Dirac$, obeying 
\beq
\mbox{\bf \{}    \bscF    \mbox{\bf ,} \, \Dirac   \mbox{\bf \}}  \speq   0   \m .
\label{C-D}
\eeq 
\n{\bf Structure 2}  The {\it space of Dirac observables} is 
\be 
\DiracObs(\lFrs)   \m .  
\ee 
\n{\bf Remark 1} The unrestricted and full notions of observables are universal over all models.

\subsection{Observables algebraic structures}\label{OAS}

\n{\bf Lemma 3} \cite{AObs} Observables themselves moreover close as further Lie brackets algebras.

\m 

\n{\u{Proof}}  Let $\sbcC$ the defining subalgebraic structure of first-class constraints for the notion of observables $\sbiO$ in question.
Our Lemma then follows from the Jacobi identity
\be
  \mbox{\bf \{} \, \bscC  \mbox{\bf ,}  \mbox{\bf \{} \, \sbiO  \mbox{\bf ,} \mbox{\bf \sbiO \}} \, \mbox{\bf \}}       \es 
- \mbox{\bf \{} \, \sbiO  \mbox{\bf ,}  \mbox{\bf \{} \, \sbiO  \mbox{\bf ,} \mbox{\bf \bscC \}} \, \mbox{\bf \}} 
- \mbox{\bf \{} \, \sbiO  \mbox{\bf ,}  \mbox{\bf \{} \, \bscC  \mbox{\bf ,} \mbox{\bf \sbiO \}} \, \mbox{\bf \}}      \speq   0     \m . \m \Box
\ee
\n{\bf Structure 1} Some physical theories moreover support further intermediate notions of classical canonical observables. 
These correspond to each closed subalgebraic structure the constraint algebraic structure possesses. 

\m 

\n{\bf Definition 1} {\it \K observables} \cite{K93} 
%
%
are quantities which form zero classical brackets with all of a given theory's first-class linear constraints,  
\beq
\mbox{\bf \{} \,    \bFlin    \mbox{\bf ,}  \, \Kuchar   \, \mbox{\bf \}}  \speq  0    \m .
\label{FLIN-K}
\eeq
Whereas $\bFlin = \bGauge$ in the more commonly encountered cases, Section \ref{CC}'s counter-examples imply the need for the following further notion.

\m 

\n{\bf Definition 2} $\lFrg$-observables alias {\it gauge-invariant quantities} 
\be 
\gauge \m \mbox{ indexed by } \m \fJ 
\ee 
obeying 
\beq
\mbox{\bf \{} \,    \bGauge    \mbox{\bf ,}  \, \gauge    \, \mbox{\bf \}}  \speq   0       \m .
\label{GAUGE-G}
\eeq
{\bf Definition 3} In cases in which $\Chronos$ self-closes, a notion of Chronos observables 
\be 
\chronos \m \mbox{ indexed by } \m \fH
\ee 
becomes meaningful, obeying 
\beq
\mbox{\bf \{} \,    \Chronos    \mbox{\bf ,}  \, \chronos    \, \mbox{\bf \}}  \speq  0 \m .
\label{CHRONOS-C}
\eeq
{\bf Structure 2} The totality of notions of observables form a 
\be
\mbox{bounded lattice } \m \lattice_{\btiO} \m \mbox{ dual to that of constraint algebras} \mma  \lattice_{\btsC}  \m .
\ee  
Its top and bottom elements are the Dirac $\biD$ and unrestricted $\unres$ notions of observables, 
whereas the middle elements are {\it A observables}, $\sbiA$ with each type of such a theory supports indexed by $\fZ$.  
Thus $\sbiO$ comprises $\unres$, $\sbiA_{\sfZ}$ and $\biD$, arranged to form $\lattice_{\btiO}$.  

\m 

\n{\bf Structure 3}  In each case closure permitting (supported by some theories but not others), the {\it space of classical Kucha\v{r} observables} is 
\be 
\KucharObs(\lFrs)  \m ,
\ee 
the {\it space of classical gauge observables} is 
\be 
\GaugeObs(\lFrs)   \m ,
\ee 
and {\it the space of Chronos observables} is 
\be 
\ChronosObs(\lFrs) \m .
\ee 
Each of these spaces, when supported, constitutes moreover an algebra. 

\m 

{\bf Structure 4} The totality of observables algebras form a 
\be
\mbox{bounded lattice } \m \lattice_{\sFrO} \m \mbox{ dual to that of constraint algebraic structures} \mma  \lattice_{\sFrC}  \m .
\ee
The algebra of unrestricted observables is the top alias unit element, and 
the algebra of Dirac        observables is the bottom alias zero element. 
All other elements are middle elements: the {\it A-observables algebraic structures}, denoted by 
\be 
\AObs(\lFrs) \m \mbox{ with each type indexed by} \m \fZ  \m . 
\ee
Thus $\Obs(\lFrS)$ comprises $\UnresObs(\lFrs)$, $\AObs(\lFrs)_{\sfZ}$ and $\DiracObs(\lFrs)$, arranged to form $\lattice_{\sFrO}$.  

\m 

\n{\bf Remark 1} Observables algebras $\Obs$, like the constraint algebraic structures $\FrC$, 
are comparable to configuration spaces $\FrQ$ and phase spaces $\Phase$ as regards the study of the the nature of Physical Law, 
and whose detailed structure is needed to understand any given theory.
This refers in particular to the topological, differential and higher-level geometric structures observables algebraic structures support, 
now with also function space and algebraic levels of structure relevant. 

\m 

\n{\bf Remark 2} This means we need to pay attention to the Tensor Calculus on observables algebraic structures as well, 
justifying our use of undertildes to keep observables-vectors distinct from contraints ones and spatial ones.  
In fact, the increased abstraction of each of these is `indexed' by my notation: no turns for spatial, one for constraints and two for observables.   

\m 

\n{\bf Remark 3} The sizes of the spaces run in the dual lattice pair run in opposition. 
I.e.\ the bigger a constraint algebraic structure, the smaller the corresponding space of observables is. 
This is clear enough from constraints acting as restrictions, adding PDEs that the observables must satisfy...

\subsection{The Problem of Observables}

The {\it Problem of Observables} -- facet 2 of the Problem of Time -- 
is that Assignment of Observables is hard, in particular for Gravitational Theory, or (partially) Background-Independent Theory more generally.
More specifically, Dirac observables   are harder to find than \K ones (see below). 
and the quantum counterparts of each are even harder to find than classical ones (see Chapter 50 of \cite{ABook}).   
The Dirac case are sufficiently hard to find for full GR that \K \cite{K93} likened strategies relying on having already obtained a full set of these 
                                                                                                            to plans involving having already caught a Unicorn.

\subsection{Simple Examples}

\n{\bf Example 1} For unconstrained theories, 
\be 
\Dirac = \Kuchar = \gauge = \unres \m .
\ee 
\n{\bf Example 2} For Minisuperspace and Spatially-Absolute Mechanics, the 
\be 
\Kuchar = \gauge = \unres
\ee 
are also trivially any quantities of the theory since these theories have no linear constraints at all.
The 
\be
\Dirac = \chronos
\ee 
are however nontrivial due to the presence of the constraint $\Chronos$.  

\m 

\n{\bf Example 3} Relational Mechanics supports distinct $\sbiK$ and $\sbiC$.

\subsection{GR Example}

\n GR's constraint subalgebraic structures support 3 notions of observables as per rows 1 and 2 of Fig \ref{C-Latt}.b).  

\m 

\n 1) Unrestricted observables $\sbiU(\bh, \bp)$. 

\m 

\n 2) Dirac observables $\sbiD(\mbox{\bf True}, \Pi^{\mbox{\scriptsize\bf True}})$, for true degrees of freedom $\mbox{\bf True}$: 
functions that commute with both $\u{\bscM}$ and $\scH$, i.e.\ with the full $\FrD\mbox{irac}(\bupSigma)$.

\m 

\n 3) Kucha\v{r} observables $\sbiK(\bG, \Pi^{\bsG})$ are supported for GR \cite{K93} as functions commuting with $\u{\bscM}$, 
a consistent notion since $\u{\bscM}$ self-closes by (\ref{Mom,Mom}) \cite{MT72}.
$\bG$ are here the 3-geometries themselves.    

\m 

\n{\bf Remark 1} Kucha\v{r}'s notion of observables does not however have theory-independent significance. 
This role is, rather, replaced by whatever non-extremal elements the bounded lattice of notions of observables of a theory happens to possess, 
i.e.\ Sec \ref{CA}'s notion of A-observables.    

\m 

\n{\bf Remark 2} See Article VI for more on GR alongside Electromagnetism and Yang--Mills Theory examples of constraints.

\subsection{Explicit solution for constrained observables} 

As regards evaluating the observables of each kind,  further formulate the bracket conditions (\ref{Shuffle-Self}) as specific PDEs.

\m

\n{\bf Structure 1} More specifically, observables' defining relations can be recast \cite{AObs2, ABook, PE-1, DO-1} 
                    for practical purposes of solution as flow PDES \cite{John, Lee2}.  

\m 
					
\n A first-order quasilinear PDE 
\be 
\sum_{A} a^{A}(x^{B}, \phi) \pa_A \phi = b(x^{B}, \phi)
\ee 
for unknown variable $\phi$ and individually-arbitrary indices $A$ and $B$ is thus recast as a flow system of ODEs 
\be 
\dot{x}^{A} = a^{A}(x^{B}, \phi)  \mma  
\ee 
\be 
\dot{\phi}       = b(x^{B}, \phi)  \m .  
\ee
Here $\dot := \pa/\pa\tau$ for some parameter $\tau$ along the flow.  

\m 

\n This is a subcase of {\bf Lie's Integral Approach to Geometrical Invariants}, uplifted moreover to a free characteristic problem for finding 
`suitably-smooth functions of phase space invariants', i.e.\ observables.  

\m 

\n{\bf Remark 1} In our current canonical setting, this amounts to an arena generalization -- to phase space -- 
of Lie's Integral Approach to Geometrical Invariants \cite{Lie, G63, PE-1, DO-1} by use of such a Flow Method.   
Demanding entire algebras rather than individual solutions to such of course involves further Lie brackets algebra level checks, 
thus also residing within Lie's Mathematics.  
Moving up the lattice moreover amounts to successive restrictions of the unreduced problem's `solution manifold' function space.  

\m 

\n{\bf Remark 2} Let us introduce the notations $\mbox{\bf --}$, $\bullet$, $\bm{\backslash}$ for functions of differences alone, dot products alone and ratios alone. 
These are of course solution spaces of observables for the simplest RPMs (see Article VII for details). 
This notation admits moreover cumulative versions such as $\bullet \bm{\backslash} \bullet$, $\mbox{\bf --}\bm{\backslash}\mbox{\bf --}$ 
and Euclidean RPM's $\mbox{\bf --} \bullet \mbox{\bf --}$, 
as well as the grand combination $\mbox{\bf --} \bullet \mbox{\bf --} \bm{\backslash}\mbox{\bf --} \bullet \mbox{\bf --}$ of similarity RPM.  
This already simplifies (II.35) and (II.42)'s notation, with some further use in Articles IV, XIII and XIV, and much further use in Article VIII.

\subsection{Expression in Terms of Observables}

{\bf Structure 1} Having an observables algebraic structure as a Function Space Thereover 
does not yet mean being able to express each physically-meaningful quantity in terms of observables.  
This involves the further step of eliminating irrelevant variables in favour of observables by whichever of Algebra and Calculus.  

\m 

\n{\bf Structure 2} Spanning, independence and bases for observables algebras (explained to be well-defined in Article XIV) 
is also a significant part of the theory of observables.
It is often convenient to pick a basis of observables, or at least a spanning set of observables, 
in terms of which all (or all required) physically-meaningful quantities can be expressed.  

\m 

\n{\bf Remark 1} Articles V and VIII set this up for RPMs and Minisuperspace, whereas Article XI extends these considerations to SIC.  
See also Article VIII for a specifically TRi Assignment of Canonical Observables.  

\m 

\n{\bf Remark 2} There are major difficulties with how observables are treated on Wikipedia \cite{W-Obs}. 
In particular, that classical {\sl constrained} theory, and classical (generally relativistic or similar) {\it spacetime}, 
each already have a nontrivial theory of such is glossed over. 
The further interaction between these sources of complexity and {\sl quantum} observables' 
own subtleties (that Wikipedia does outline) is moreover left out altogether. 
These points, and consequent distinctions in definitions of types of observables are due review.  
Articles VII and VIII indicate, among other things, how to bring more subtlety, truth and clarity to Wikipedia's article on observables, 
in the constrained-canonical and spacetime settings respectively.   
The current state of the Wikipedia page on the Problem of Time is more generally commented upon in Article XIII. 

\m 

\n{\bf Remark 3} With \cite{AObs, ABook, PE-1, DO-1} and the current Series, the days of finding individual or few observables are over. 
Solutions to the Problem of Observables are to involve, rather, 
a whole 'function space that is also an algebraic structure' of these per theory per notion of observables consistently supported by that theory. 
In other words, given a theory, Kucha\v{r}'s quest to find a Unicorn for it needs to be extended to finding that theory's entire `Unicornucopia' 
(space of physical observables). 
Or at least Unicornucopia enough to be able to complete Expression in Terms of Observables for every physical proposition contained by that theory, 
a matter to which we return in Article VIII.

\section{Spacetime Constructability (Aspect A)}\label{SpC} 

\subsection{Motivation}

This an independent development to Assignment of Observables, as indicated by the fork in Fig 7. 
Though if new consistent theories emerge in this way, one then needs to plug each of these into the observables-determining procedure. 

\m 

\n Suppose one assumes less structure than is present in GR's notion of spacetime, one needs to recover it from what structure is assumed (at least in a suitable limit).  
This can be a hard venture; in particular, the less structure is assumed, the harder it is.
This aspect was originally known as `Spacetime Reconstruction', the Author however takes this name to be too steeped in assuming spacetime primality. 
We thus use instead the terms `Spacetime Constructability'     for the Background Independence aspect,  
                              `Spacetime Construction Problem' for the Problem of Time facet in cases in which this is blocked, and 
                              `Spacetime Construction'         for corresponding strategies.  
Some quantum-level motivation for this due to Wheeler \cite{Battelle} is outlined in Sec IV.7.

\subsection{Rigidity in Feeding Deformed Families through the Dirac Algorithm}

\n A second answer to Wheeler's question (II.25) is provided by sense A). 

\m 

\n This goes beyond the first answer -- Hojman, \K and Teitelboim's Deformation Approach \cite{HKT} -- 
which assumes embeddability into spacetime (see Article XII for more details). 

\m 

\n The second answer -- Barbour, Foster, \'{o} Murchadha and the Author's Relational Approach \cite{RWR, Phan, AM13, ABook} -- however, 
goes further by not assuming spacetime structure. 
Since this approach recovers spacetime, moreover, a Spacetime Construction has arisen. 

\m 

\n{\bf Modelling assumptions} In this approach
\be 
\mbox{3-spaces } \m \bigupsigma
\ee 
are assumed in place of hypersurfaces $\bupSigma$ {\sl in spacetime}. 
One also assumes Temporal and Configurational Relationalism first principles.
One can moreover proceed with just a BSW-type action rather than a Temporally Relational one, 
i.e.\ for now in isolation of that Background Independence aspect.  

\m

\n{\bf Theorem 1 (Spacetime Construction)} Consider the general family of relational actions  
\be 
{\cal S}_{\st\sr\si\sa\sll}^{w,y,a,b}  \:=  \iint_{\bigupsigma} \d^3 x \sqrt{\overline{a \, {\cal R} + b}} \, \d \ms_{w,y} \m . 
\label{trial} 
\ee
Here, $a$, $b$ are constants, overline denotes densitization, and $\d \ms_{w,y}$ is built out of the usual shift $\u{\upbeta}$ 
and the more general -- if still ultralocal -- supermetric 
\be 
\bM_{w,y} \m  \mbox{ with components } \m  
\mM^{abcd}_{w,y}  \:=  \frac{\sqrt{\mh}}{y}\{\mh^{ac}\mh^{bd} - w \, \mh^{ab}\mh^{cd}\}  \m .
\ee  
This ansatz also assumes that the kinetic metric is homogeneous quadratic in the changes.   

\m 

\n $\bM_{w,y}$'s inverse 
\be 
\bN^{x,y} \m \mbox{ has components } \m  
\mN_{abcd}^{x,y}  \:=  \frac{y}{\sqrt{h}}\left\{  \mh_{ac}\mh_{bd} -  \frac{x}{2} \,\mh_{ab}\mh_{cd} \right\}
\ee 
for  
\be 
x \:= \frac{2 \, w}{ 3 \, w - 1 } \m . 
\ee 
The parametrization by $x$ has been chosen such that GR is the $w = 1 = x$ case. 

\m 

\n The $\Chronos$ constraints arising from this as per Dirac's argument form the more general family of 
\be 
\scH_{\st\sr\si\sa\sll}  \es  \scH_{x,y,a,b}  
                         \:=  {||\bp||_{\sbiN}^{x, y}}^2 - \overline{a \, {\cal R} + b}  
						 \es  0                                                                 \m .
\ee 
The $\Chronos$ self-bracket now picks up an extra additive term as an {\it obstruction term to having a brackets algebraic structure}. 
\be
2 \times a \times y \times \{ 1 - x \} \times (    \bcalD \, \mp \, | \, \xi \, \overleftrightarrow{\bpa} \zeta    ) \m .  
\label{4-Factors}
\ee 
This has four factors, each of which being zero providing a different way in which to attain consistency. 
GR follows from the third of these setting $x = 1$ \cite{RWR}.
 
\m 

\n{\bf Remark 1} When working with minimally-coupled matter terms as well, 
the above obstruction term is accompanied by a second obstruction term \cite{RWR, AM13, ABook} (see Article IX for details).
Its factors give the ambiguity Einstein faced of whether the universal Relativity is locally Galilean or Lorentzian. 
This ambiguity is based on whether the fundamental universal propagation speed $c$ -- often referred to as `speed of light' -- takes an infinite value, 
or a fixed finite value. 
(Carrollian Relativity, corresponding to zero such speed, now features as a third option as well.) 
This ambiguity is now moreover realized in the {\sl explicit mathematical form} of a string of numerical factors 
in what would otherwise be an {\sl obstruction term to having a brackets algebraic structure of constraints}.  
In particular, the GR spacetime solution to the first obstruction term is now accompanied by the condition 
from the second obstruction term that {\sl the locally-Lorentzian relativity of SR is obligatory}.  
This can be viewed as all minimally-coupled matter sharing the same null cone because each matter field is separately obliged to share {\sl Gravity}'s null cone.  

\m 

\n{\bf Remark 2} This string of numerical factors moreover arises {\sl from the Dirac Algorithm} as the choice of factors among which one needs 
to vanish in order to avoid the constraint algebroid picking up an obstruction term.
This is substantially different from the form of Einstein's dichotomy between universal local Galilean or Lorentzian Relativity!  
Now mere algebra {\sl gives us} this dichotomy (within a greater fork: 4-way), rather than it taking an Einstein to intuit... 

\m 

\n{\bf Remark 3} The first factor in (\ref{4-Factors}) vanishing corresponds to Strong Gravity: a natural setting for Carrollian Relativity. 
The second factor corresponds to Galileo--Riemann Geometrostatics, features in the consistent brackets algebraic structure context in e.g.\ \cite{Phan}.  
The fourth, now merely weakly vanishing factor, corresponds to a privileged constant mean curvature (CMC) foliation.
For this, the GR constraints decouple: methodology already well-established in the GR Initial-Value Problem \cite{York73}.  
The combination of GR's particular $\scH$ alongside local Lorentzian Relativity and 
embeddability into GR spacetime thus arises as one of very few consistent possibilities, the others of which are at least of conceptual interest.  

\m 

\n{\bf Remark 4} See Article IX for a specifically TRi Constructability of Spacetime.

\subsection{Simpler examples}

\n{\bf Example 1)} Newtonian Mechanics and RPMs have no notion of Spacetime Constructability, due to these having no notion of spacetime in the first place.  
The analogue of this Section's working for these gives that their simpler constraint brackets exhibit no corresponding obstruction terms.

\m 

\n{\bf Example 2)} Minisuperspace modelled within the foliation by hypersurfaces privileged by homogeneity. 
This greatly simplifies the study of spacetime.
In this case, moreover, the obstruction term vanishes, since it contains a spatial derivative operator factor $\bcalD$ that now acts upon a homogeneous entity.
Minisuperspace additionally has no $\u{\bscM}$. 
The constraint algebraic structure consequently ends up being just $\scH_{\st\sr\si\sa\sll}$ commuting with itself (for any $a$, $b$, $w$, $y$). 
Thus generalized Minisuperspace is {\sl not} restricted by constraint algebraic structure consistency.
This is very similar in content to the metrodynamical case of Strong Gravity. 
[In the former case, ${\cal R}$ is just a spatial constant, so $a \, {\cal R} + b$ behaves just like $0 + b^{\prime}$ for a new constant $b^{\prime}$.]

\section{Constructability of Space from less Space Structure Assumed}\label{CMG}
%

The following 2-parameter family ansatz gives the general (bosonic vectorial) quadratic generator in $\geq 2$-$d$: 
\be 
\u{Q}^{\st\sr\si\sa\sll}_{\mu, \nu}  \:=  \mu \, ||x||^2 \u{\pa} + \nu \, \u{x} \, (\u{x} \cdot \u{\nabla} )                             \m .  
\ee 
This follows from considering the general fourth-order isotropic tensor contracted into a symmetric object $x^Ax^B$.  

\m 

\n{\bf Theorem 2 (Flat Space Top Geometry)} \cite{A-Brackets} 
For $d \geq 2$, $\u{Q}^{\st\sr\si\sa\sll}_{\mu, \nu}$ self-closes only if $\mu \, (2 \, \mu + \nu) = 0$.

\m 

\n{\bf Remark 1} This arises as a strong vanishing to avoid an obstructory cofactor; 
it amounts to Constructability of more structured spatial geometry from less structured spatial geometry. 

\m 

\n {\bf Remark 2} The first factor vanishing is a recovery of the special-projective generator 
\be 
\u{Q} =  \u{x} \, (\u{x} \cdot \u{\nabla} )            \m , 
\ee 
whereas the second is a recovery of the special-conformal generator            
\be 
\u{K} =  ||x||^2 \u{\pa} - 2 \, \u{x} \, (\u{x} \cdot \u{\nabla})  \m . 
\ee 
In this manner, the Conformal Geometry versus Projective Geometry alternative for flat-space `top geometry' is recovered.  			   
{\it Q.E.D.} that Lie Algebraic Rigidity  
outside of Dirac Rigidity in the Poisson brackets subcase, is a realized phenomenon.

\section{Spacetime's own Relationalism (Aspect 0$^{\prime}$)}\label{SpR} 

Either having constructed GR spacetime $\Frm$, or starting afresh with spacetime primality, 
one needs to take into account that spacetime has a Relationalism of its own: Spacetime Relationalism.

\m 

\n{\bf Spacetime Relationalism i)} There are no background spacetime structures; in particular there are no indefinite-signature background spacetime metrics.  
Fixed background spacetime metrics are also more well-known than fixed background space metrics.  

\m 

\n{\bf Spacetime Relationalism ii)} Consider not just a spacetime manifold $\Frm$ but also a $\lFrg_{\sS}$ 
of transformations acting upon $\Frm$ that are taken to be physically redundant.

\m

\n In GR-like theories, this usually taken to involve as spacetime automorphism group 
\be 
\lFrg_{\sS} = Diff(\Frm)  \m :
\label{Diff-M}
\ee 
the (unsplit) spacetime 4-diffeomorphisms (or possibly some generalization).  

\m 

\n{\bf Remark 1} $\Frm$ can moreover additionally be equipped with matter fields in addition to the metric. 

\m 

\n{\bf Remark 2} Spacetime Relationalism i) can then be extended to include no background internal structures associated with spacetime.
Note the difference between these and structure on spacetime, in direct parallel to the stated parts of Configuational Relationalism.

\m

\n{\bf Remark 3} Spacetime Relationalism ii)'s $\lFrg_{\sS}$ can furthermore have a part acting internally on a subset of the fields. 
We comment further on this in Article X.

\section{Spacetime Generator Closure (Aspect 1$^{\prime}$)}\label{Sp-GC}

\subsection{Closure more generally}

Questions of Closure follow on from having a set of generators due to Relationalism acting as a Generator Provider. 

\m 

\n For a wide range of situaions, 
Closure is to be assessed by forming Lie brackets \cite{Stillwell-Lie, Serre-Lie}\footnote{See Chapter 24 of \cite{ABook} and \cite{Nambu} for further generalizations
of relevance to e.g.\ Supergravity or M-Theory, and \cite{Nijenhuis} for other extensions.}
\be 
\mbox{\bf |[} \m \mbox{\bf ,} \, \m \mbox{\bf ]|}
\ee 
which are antisymmetric, bilinear and obey the {\it Jacobi identity} 
\be 
\mbox{\bf |[} \, \mbox{\bf |[} A \mbox{\bf ,} \, B \mbox{\bf ]|} \mbox{\bf ,} \, C \m \mbox{\bf ]|} + cycles = 0   \m . 
\ee 
These now act on objects which may not necessarily be constraints; 
one way these can be more general is as a continuous group's infinitesimal generators.
Another way involves constraints-and-specifiers. 
As Article XIV explains, these two ways can furthermore combine to the case of generators-and-specifiers.  

\m 

\n One idea is then that a set of whichever of the preceding may imply further such under Lie brackets, 
or may, rather (and perhaps eventually) close. 

\m 

\n{\bf Structural extension} While strongly vanishing brackets are clearly universal, 
there is moreover some motivation to extend Dirac's notion of weakly vanishing 
from Poisson or Dirac brackets of constraints to general Lie brackets of generators, as follows.

\m 

\n{\bf Naming 1} We coin {\it generator-weakly vanishing} to mean vanishing up to linear combinations of the generators. 
We use {\it Lie-weakly} as an honorific synonym, at least in the present context 
(see Article VII and \cite{Nijenhuis, Nambu} for generators of continuous algebraic structures more general than Lie's). 

\m 

\n This is to be contrasted with prior use of `by which' to be replaced with {\it constraints-weakly}, for which {\it Dirac-weakly} is an actual historical synonym. 
This furthermore extends to subnotions of Dirac weakness, to be named by preceding `constraints-' with which kind.

\m 

\n Let us thus extend our weak equality notation $\approx$ from Dirac to general Lie setting, and our portmanteau equality notation $\peq$ as well. 

\m 

\n What we are beginning to describe above is in effect a {\it Lie Algorithm} extension of the notion of {\it Dirac Algorithm}; 
this is developed in detail in Article XIV.

\subsection{Example of GR spacetime}

\n GR Spacetime also possesses its own version of Closure -- now not Constraint Closure but Generator Closure, in particular for  
GR spacetime diffeomorphisms \ref{Diff-M} whose generators obey  
\be
\mbox{\bf |[} \, (  \, \vec{\bcalD} \, | \, \vec{\bX}) \mbox{\bf ,} \, (    \vec{\bcalD} \,  | \,  \vec{\bY}  \,  )  \, \mbox{\bf ]|} 
         \es     (  \, \vec{\bcalD} \, | \, \stackrel{\longrightarrow}{\mbox{[ \, \bX, \,  \bY \, ]}  }    )                                                                          \m .  
\label{Lie-2}
\ee
$\vec{\bX}$ and $\vec{\bY}$ are here spacetime smearing variables, whereas $[ \m , \m ]$ is Differential Geometry's commutator of two vectors.   
This equation is a subcase of generator-weakly vanishing Generator Closure.   

\m 

\n{\bf Remark 1} Classical GR's $Diff(\bupSigma)$ closes to form a Lie algebra, in parallel to how $Diff(\bupSigma)$ does.  

\m 

\n{\bf Remark 2} $Diff(\Frm)$ also shares further specific features with $Diff(\bupSigma)$, such as its right hand side being of Lie derivative form.

\m 

\n{\bf Remark 3} {\it All three kinds of Relationalism considered up to this point are thus implemented by Lie derivatives.}  

\m

\n{\bf Remark 4} Some differences are that whereas the generators of $Diff(\bupSigma)$ are conventionally associated with dynamical constraints, 
those of $Diff(\Frm)$ are not. 
$Diff(\bupSigma)$'s -- but not $Diff(\Frm)$'s -- Lie bracket is moreover conventionally taken to be a Poisson bracket.  

\m 

\n{\bf Remark 5} This is to be contrasted this with split space-time's Dirac algebroid $\FrD\mbox{irac}(\bupSigma)$, 
which Sec 10 explains is much larger.  
%
{\begin{figure}[!ht]
\centering
\includegraphics[width=0.45\textwidth]{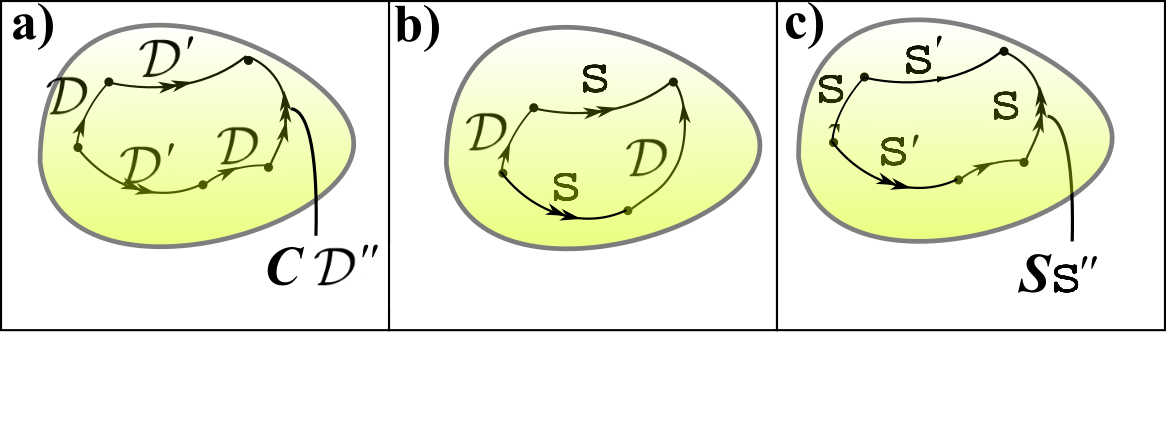}
\caption[Text der im Bilderverzeichnis auftaucht]{\footnotesize{a) Spacetime diffeomorphisms close as a Lie algebra. 

\m 

\n b) The (strong case of) spacetime observables condition. 

\m 

\n c) Spacetime observables themselves close as an algebraic structure. } } 
\label{STR-Figure}\end{figure}} 

\subsection{Generator Closure Problem}

Each enumerated Constraint Closure Problem has a corresponding Generator Closure Problem (other than breakdown of Temporal versus Configurational distinction).  
The current Series, moreover, does no involve any $\lFrg_{\sS}$ that produces specifier equations, or second-classness, 
limiting us to the following subset of counterparts of Constraint Closure Problems. 

\m 

\n{\bf Generator Closure Problem by Sufficient-Cascade Inconsistency}. 

\m 

\n{\bf Enforced-Group-Extension Generator Closure Problem}. 

\m 

\n{\bf Topologically-Rooted Generator Closure Problem}. 
%
%

\m 

\n{\bf Remark 1} Spacetime generator subalgebraic strucures are also possible (see Article X for examples) 
and the totality of them form a lattice $\lattice_{\sFrG}$.

\section{Spacetime Observables (Aspect 2$^{\prime}$)}\label{Sp-Obs-Sec} 

\n{\bf Example 1} Unrestricted spacetime observables 
\be 
\bttu 
\ee
forming 
\be 
\UnresObs(\lFrs) = \FrC^{\infty}(\FrM)  \m . 
\ee 
{\bf Structure 1} We usually however consider entities which commute with the spacetime symmetry group $\lFrg_{\sS}$'s generators: 
{\it fully restricted spacetime observables}, $\bttf$.

\m 

\n{\bf Example 2} In particular, one can place functions over spacetime which are $Diff(\Frm)$-invariant, i.e.\ Lie-brackets commutants with with $Diff(\Frm)$'s generators, 
\be
\mbox{\bf |[} \, (  \,  \bcalD \, | \,  \bY   \,  ) \,  \mbox{\bf ,} \, (  \,  \btts \, | \, \bZ  \,  ) \, \mbox{\bf ]|}  \es 0  \m 
\label{Sp-Obs} 
\ee
for spacetime smearing variables $\bY$ and $\bZ$, or perhaps the generator-weak equality extension of this equation.  

\m 

\n{\bf Structure 2} These $\bif$ form an infinite-$d$ Lie algebra
\be 
\DiffObs(\lFrs)  \m .
\ee 
$\bis$ more generally form a dual lattice.

\section{Spacetime Constructed from less structured spacetime (Aspect 3$^{\prime}$)}\label{Spt-Cons-Sec} 

\n{\bf Remark 1} Sec \ref{CMG}'s working carries over to the indefinite case of flat space geometry (we spell this out in Article X).   

\m 

\n Thereby, more structured spacetime geometry is demonstrated to be Constructible from less structured spacetime geometry. 

\m 

\n{\bf Remark 2} While space from less structured space has no temporal content to it, spacetime from less structured spacetime does.
Space from less structured space is thus a Background Independence aspect 3 that is not a Problem of Time facet, 
whereas spacetime from less structured spacetime is a further facet of the Problem of Time new to the current Article.  

\m 

\n{\bf Remark 3} Beyond some point in levels of structure, one might cease to call it `spacetime', 
not least because of loss of spatio-temporal (and thus in part temporal) features. 
Relative to space, it is still a `larger' state space, for a while characterized by having dimension higher by 1.  
 
 
\section{Foliation Independence (Aspect B)}\label{Fol} 
%
{            \begin{figure}[!ht]
\centering
\includegraphics[width=1.0\textwidth]{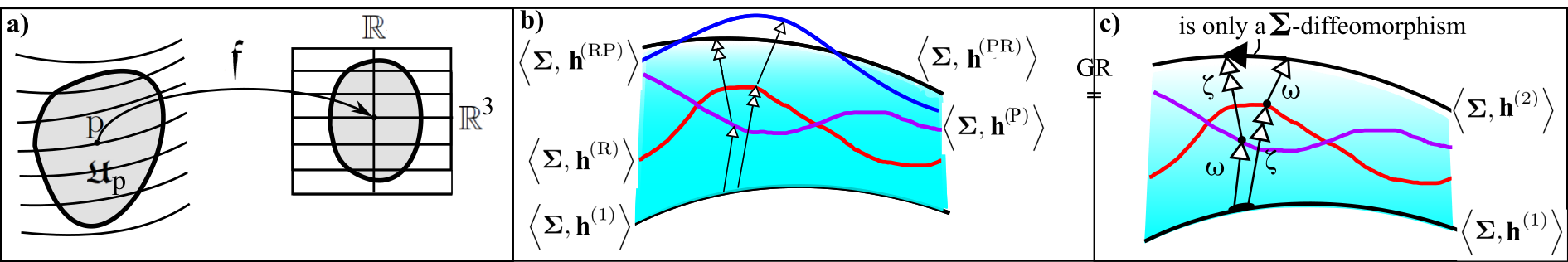}
\caption[Text der im Bilderverzeichnis auftaucht]{ \footnotesize{
a)Illustrating the nature of {\it foliation} $\sFrf$: a decorated, or, more precisely, rigged \cite{Lee2} version of the standard Differential-Geometric definition of chart. 

\m 

\n b) Posing Refoliation Invariance: is going from spatial hypersurfaces (1) to (2) 
via the red (R) intermediary hypersurface being physically the same as going via the purple (P) intermediary surface? 
If so, the blue and black hypersurfaces would coincide. 

\m 

\n c) For GR, however, (\ref{Ham,Ham}) gives that the black and blue hypersurfaces can at most differ by a spatial diffeomorphism, 
and so must coincide as the same geometrical entity (PR = 2 = RP)
This can be seen as the diffeomorphism establishing that an a priori distinct square of maps is in fact commutative.} }
\label{Refol-Inv}\end{figure}            }

\n{\bf Remark 1} GR spacetime admits multiple foliations.    
At least at first sight, this property is lost in the geometrodynamical formulation. 

\m

\n{\bf Foliation Independence} is a privileged coordinate independence aspect that is desirable on Generally-Relativistic grounds.

\m 
 
\n{\bf Foliation Dependence Problem} is the corresponding Problem of Time facet, 
whether due to a model arena not exhibiting the aspect or it remaining undemonstrated whether it does.   
Foliation Dependence runs against the basic principles that GR contributes to Physics.

\m 

\n{\bf Remark 2} Foliation (In)Dependence clearly involves time since each foliation by spacelike hypersurfaces is dual to a GR timefunction.   

\m 

\n{\bf Strategy B-1} We firstly set this up by formulating the corresponding foliation kinematics along Isham's lines \cite{I93}. 

\m 

\n A new TRi version (rather than the original ADM-like \cite{ADM} version) of this kinematics 
is again required to retain compatibility with Temporal Relationalism: {\it TRiFol} \cite{TRiFol} (`Fol' standing for foliations). 
This is provided in Article XII.  

\m 

\n{\bf Strategy B-2} {\it Refoliation Invariance} is encapsulated by evolving via each of Fig \ref{Refol-Inv}.b)'s red and purple hypersurfaces 
giving the same physical answer as regards the final hypersurface. 

\m 

\n{\bf Remark 3} So whereas Foliation Independence is a matter of freedom in how to strut spatial hypersurfaces together, 
Refoliation Invariance instead concerns passing between such struttings.  
An arena possessing Refoliation Invariance is a victory condition as regards overcoming the Foliation Dependence Problem.

\m 

\n{\bf Theorem 3 (Teitelboim)} 
The space--time split of GR spacetime is Refoliation Invariant \cite{Tei73}.

\m 

\n{\bf Remark 4} GR spacetime is thus not just a single strutting together of spaces like Newtonian space-time is.  
GR spacetime manages, rather, to be many such struttings at once in a physically mutually consistent manner, as per Fig \ref{Refol-Inv}.c).
Indeed, this is how GR is able to encode consistently the experiences of fleets of observers moving in whichever way they please. 

\m 

\n{\u{Proof}} Refoliation Invariance compares triples of hypersurfaces.

\m 

\n In each triple, one starts from the same hypersurface and subsequently applies the same two operations, but in opposite orders in each case.

\m 

\n The question is then whether the outcome of these two different orders is the same 
[Fig \ref{CC-Figure}.e) as further explained in Fig \ref{Refol-Inv}.b)]. 

\m 

\n This is moreover in direct correspondence with Fig \ref{CC-Figure}.f)'s commuting pentagon \cite{Higher-Lie}. 

\m 

\n Since the individual operations involved are actions of $\scH$, one is led to the commutator of two $\scH$'s.  

\m 

\n Then indeed, as Teitelboim pointed out \cite{Tei73}, 
the form of this part (\ref{Ham,Ham}) of GR's Dirac constraint algebroid guarantees Refoliation Invariance.   

\m 

\n This is achieved by the two end hypersurfaces coinciding up to a diffeomorphism of that hypersurface, 
as per the right hand side of (\ref{Ham,Ham}).  

\m 

\n The third equation in GR's Dirac algebroid (\ref{Ham,Ham}) 
is thus none other than a local algebraic formulation of Refoliation Invariance (Fig \ref{Refol-Inv}.c). $\Box$  

\m 

\n{\bf Remark 5} Theorem 3 also lies within Lie's Mathematics \cite{Higher-Lie}. 
Though various subsequent arena developments are required to frame this application: 
work of Dirac \cite{Dirac}, Teitelboim \cite{Tei73}, Kucha\v{r} \cite{Bubble} and some further formulations from the late 1970s reviewed within \cite{I93, ABook}.

\m 

\n{\bf Remark 6} Resolving Foliation Independence by Refoliation Invariance {\sl requires} 
the enlarged structure of an algebroid to keep track of all the foliations. 
%
%
This is tied to 
\be 
\FrD\mbox{irac}(\bupSigma) = Diff(\FrM, Fol) \m >> \m Diff(\Frm) 
\ee
with $Fol$ entering as a variable, and the vast difference in size is due to the large freedom in how to foliate $\FrM$.  
This vastness parallels how functions over a space can have vastly larger variety than constants, 
indeed being tied to the former having structure functions in place of structure constants.
This enlarged structure and keeping track moreover models the experience of each possible fleet of observers navigating through the spacetime; 
see Article XII for further details.

\m 

\n{\bf Remark 7} An underlying explanation of the phenomenon of Refoliation Invariance in GR-as-Geometrodynamics is that 
it has the status of a {\sl hidden} symmetry encoded by $\scH$.  

\m

\n{\bf Remark 8} While Theorem 3 was not originally formulated in a TRi manner, this can be mitigated (see Article XII or \cite{TRiFol, ABook}).

\section{Conclusion}\label{Conclusion} 

\subsection{Ordering aspects or facets}
%
 {            \begin{figure}[!ht]
 \centering
 \includegraphics[width=0.85\textwidth]{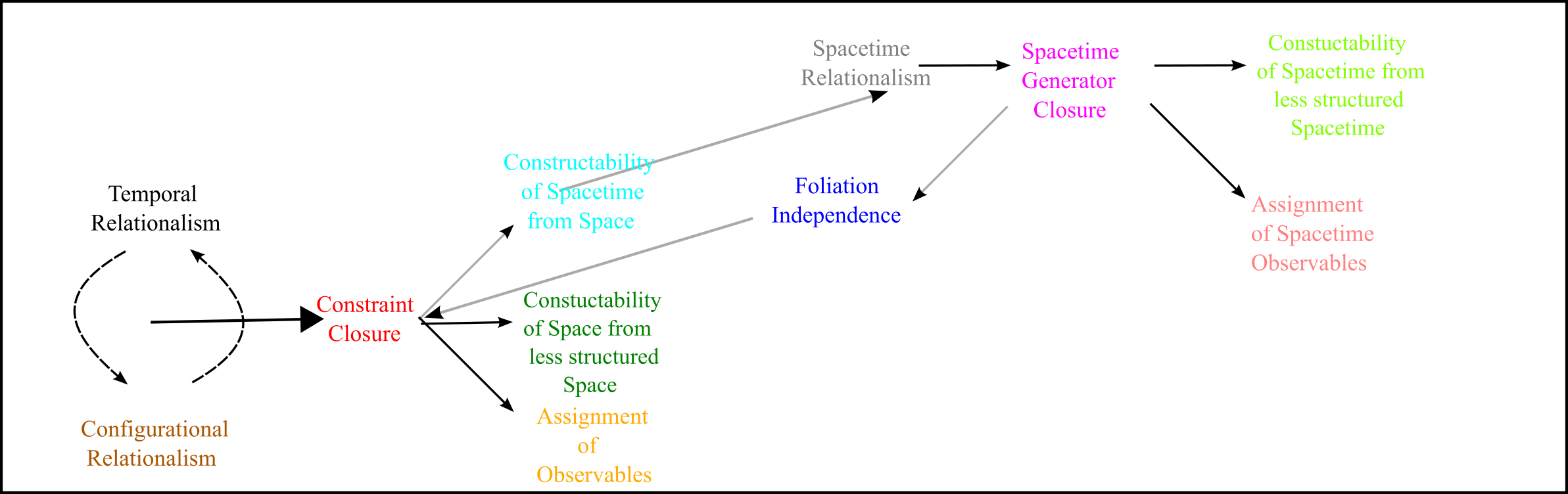} 
 \caption[Text der im Bilderverzeichnis auftaucht]{\footnotesize{Classical order of incorporation of Background Independence aspects, 
i.e.\ of overcoming corresponding Problem of Time facets.} } 
 \label{New-Gates} \end{figure}          }

\n Fig \ref{New-Gates} places the local classical aspects of Background Independence in order of incorporation. 
This non-linear order can be viewed as one in which the corresponding Problem of Time facets can be overcome, 
or as a logical order in which Backgound Independence's aspects can be consistently built up.  
This answers a question of \Ks \cite{K93}, for now at the classical level. 
The conceptual and technical reasons for this ordering succeeding, moreover, are almost everything we need to order the quantum-level aspects or facets as well.

\m 
 
\n Further commentary on this, and the end-summary progression from facets to aspects awaits Article IV's quantum treatment, 
after which Article IV concludes on joint behalf of the piecemeal-facet Articles I to IV.  



\end{document}